\documentclass[sigconf]{acmart}
\AtBeginDocument{%
  \providecommand\BibTeX{{%
    \normalfont B\kern-0.5em{\scshape i\kern-0.25em b}\kern-0.8em\TeX}}}

\copyrightyear{2025}
\acmYear{2025}
\setcopyright{cc}
\setcctype{by}
\acmConference[IUI '25]{30th International Conference on Intelligent User Interfaces}{March 24--27, 2025}{Cagliari, Italy}
\acmBooktitle{30th International Conference on Intelligent User Interfaces (IUI '25), March 24--27, 2025, Cagliari, Italy}
\acmDOI{10.1145/3708359.3712133}
\acmISBN{979-8-4007-1306-4/25/03}

\usepackage{xcolor, colortbl}
\usepackage{xspace}
\usepackage{multirow}
\usepackage{subfigure}
\usepackage{enumitem}
\usepackage{svg}
\usepackage{framed}

\newcommand{\control}{\textbf{\texttt{Control}}}
\newcommand{\chatxai}{\textbf{\texttt{CXAI}}}
\newcommand{\chatpersonalized}{\textbf{\texttt{ECXAI}}}
\newcommand{\dashboard}{\textbf{\texttt{Dashboard}}}
\newcommand{\chatagent}{\textbf{\texttt{LLM Agent}}}

\newcommand{\reviewer}[1]{}
\newcommand{\glcomment}[1]{}

\newcommand{\ug}[1]{}

\newcommand{\revise}[1]{{#1}}
\newcommand{\ie}{\textit{i.e.,}~}
\newcommand{\eg}{\textit{e.g.,}~}
\newcommand{\etal}{\emph{et al.}\xspace}
\newcommand{\paratitle}[1]{\vspace{1.0ex}\noindent\textbf{#1}}

\begin{document}

\title{Is Conversational XAI All You Need? Human-AI Decision Making With a Conversational XAI Assistant}

\author{Gaole He}
\affiliation{%
  \institution{Delft University of Technology}
  \city{Delft}
  \country{The Netherlands}}
\email{g.he@tudelft.nl}

\author{Nilay Aishwarya}
\affiliation{%
  \institution{Delft University of Technology}
  \city{Delft}
  \country{The Netherlands}
}
\email{nilayaishwarya02@gmail.com}

\author{Ujwal Gadiraju}
\affiliation{%
  \institution{Delft University of Technology}
  \city{Delft}
  \country{The Netherlands}
}
\email{u.k.gadiraju@tudelft.nl}
\renewcommand{\shorttitle}{Human-AI Decision Making With a Conversational XAI Assistant}
\begin{abstract}
  Explainable artificial intelligence (XAI) methods are being proposed to help interpret and understand how AI systems reach specific predictions. Inspired by prior work on conversational user interfaces, we argue that augmenting existing XAI methods with conversational user interfaces can increase user engagement and boost user understanding of the AI system. 
  In this paper, we explored the impact of 
  a conversational XAI interface on users' understanding of the AI system, their trust, and reliance on the AI system. 
  In comparison to an XAI dashboard, we found that the conversational XAI interface can bring about 
  a better understanding of the AI system among users and higher user trust. 
  However, users of both the XAI dashboard and conversational XAI interfaces showed clear over-reliance on the AI system. 
  Enhanced conversations powered by large language model (LLM) agents amplified over-reliance. 
  Based on our findings, we reason that the potential cause of such over-reliance is the illusion of explanatory depth that is concomitant with both XAI interfaces. 
  Our findings have important implications for designing effective conversational XAI interfaces to facilitate appropriate reliance and improve human-AI collaboration. 
  
\end{abstract}





\maketitle

\section{Introduction}


In recent years, deep learning-based AI systems have brought about tremendous possibilities to change and affect our daily life~\cite{parloff2016deep,sejnowski2018deep}. 
Due to the intrinsic opaqueness of such systems, {automating critical decision making by using AI systems is far from reliable} 
~\cite{ehsan2022seamful}. 
{However, leveraging such powerful AI systems to \textit{assist} and \textit{empower} human decision makers is an alternative that has gained prominence~\cite{lai2021towards}.} 
In such a collaborative decision making process, explanations are incorporated to increase intelligibility and ensure that decision makers can make informed decisions~\cite{de2020artificial}.
Post-hoc explainable AI (XAI) methods are typically used to help explain AI predictions from deep learning-based AI systems. 


{To realize the goal of complementary team performance, users of an AI system are expected to rely appropriately on AI advice~\cite{schemmer2022should}. 
Such appropriate reliance requires a comprehensive understanding of the AI system and its underlying rationale alongside the AI advice~\cite{lee2004trust,bussone2015role,schemmer2023appropriate}, which play important roles in calibrating user trust and reliance behaviors~\cite{Zhang-FAT-2020,vasconcelos2023explanations}. According to several empirical studies in human-AI collaboration~\cite{Zhang-FAT-2020,lai2021towards,wang2021explanations}, most XAI methods are not as helpful as expected and are even harmful at times (\eg causing over-reliance). 
The reasons behind this are multi-fold: (1) Most existing XAI methods can only provide specific types of information~\cite{liao2020questioning} (\eg feature importance~\cite{lundberg2017unified}, counterfactual reasoning~\cite{yang2022mace}). (2) 
In practice, there are diverse stakeholders of AI systems~\cite{preece2018stakeholders,langer2021we} (\eg developers, experts, and laypeople) 
{having different levels of domain expertise and AI literacy}. 
(3) 
The information needs of diverse stakeholders can vary greatly. 
Thus, a specific type of XAI method can seldom address varying information needs, resulting in 
a lack of understanding of the AI system.}

{Based on folk concepts in the theory of mind literature, Jacovi \etal~\cite{jacovi2023diagnosing} argue that successful explanations can provide users with the necessary components  to build a coherent mental model. 
We extrapolate that to make critical decisions with AI assistance, users need to build a relatively more complete and coherent mental model by exploring different explanations provided by XAI methods.
However, such a process can be complex---it requires processing information based on a variety of aspects, depending on the XAI methods. 
When presenting tailored explanations for specific audiences, designers need to trade off the simplicity and completeness of the explanations~\cite{hohman2019gamut}. 
Instead of selecting a single specific explanation, an XAI dashboard enables users to explore their information needs by providing them access to their 
desired explanations on demand. 
Such an interactive interface can bring forth the advantages of both simplicity and completeness and has been increasingly recognized as an effective design 
~\cite{nori2019interpretml,wenzhuo2022-omnixai}.
However, not all users have the necessary AI knowledge and experience to understand or benefit from such explanations~\cite{liao2020questioning}. 
Nor can all users articulate their information needs and find suitable XAI methods to address their concerns~\cite{slack2023explaining}. 
Therefore, we need a more flexible, dynamic, and personalized approach to resolving users' explanation needs.}

Conversational user interfaces can provide a human-like interaction~\cite{moore2017conversational} and simplify complex tasks with filtered information~\cite{brennan1990conversation}, which can bring 
better user experience and higher user engagement. {Inspired by prior work on conversational user interfaces for XAI~\cite{slack2023explaining}, we argue that augmenting existing XAI methods with conversational interfaces can potentially boost users' understanding of the AI system through an improved exploration of their explanation needs. 
{
Such interaction may benefit humans by fostering increased engagement and helping build a relatively more coherent and complete mental model 
that aids their information needs. 
Thus far, only a few studies~\cite{slack2022talktomodel,lakkaraju2022rethinking,wijekoon2022behaviour,shen2023convxai,mindlin2024measuring} 
have explored how conversational interfaces can be combined with XAI methods. 
However, existing work has not systematically explored the impact of conversational XAI interfaces on user trust and reliance {in the context of critical decision making}. 
Our work presents a 
study that addresses this under-explored research and empirical gap.
}

In this paper, we explored how 
conversational XAI interfaces 
shape user understanding of an AI system. 
To this end, we aim to address the following research questions:
\begin{framed}
\begin{small}
\noindent\textbf{RQ1}: \textit{How does a 
conversational XAI interface shape user understanding of an AI system\revise{, in comparison with the XAI Dashboard}?}\\
\textbf{RQ2}: \textit{How does 
a conversational XAI interface influence user trust and reliance on an AI system\revise{, in comparison with the XAI Dashboard}?}
\end{small}
\end{framed}
To answer these questions, we conducted an empirical study ($N = \revise{306}$), exploring human-AI collaborative decision making in a loan approval task (\ie making a binary decision based on a loan applicant profile). 
{To further our understanding of the impact of {enhanced conversation with flexible user input and high-quality text responses based on XAI outcomes}, \revise{we considered large language model (LLM) agents to power the conversational XAI interface.}
Overall, we found that 
users with {conversational} XAI interfaces 
tended to rely more on the AI system. 
However, such increased reliance did not always translate into appropriate reliance. 
Instead, it was characterized by 
clear patterns of over-reliance.  
Compared to an XAI dashboard, we observed limited improvements in user understanding and trust brought forth by the conversational XAI interface.} 
{We found a strong correlation between most measures of user understanding and user trust with users' reliance behaviors. 

Our results collectively suggest that both the XAI dashboard and the conversational XAI interface worked as persuasive technology. 
{Leveraging LLM agents to power the conversational interface can increase the perceived plausibility of explanations, potentially amplifying such impact.}
}
%
These observations highlight that supporting specific AI advice with interactive XAI interfaces can lead to creating an illusion of explanatory depth. 
To this end, users may overestimate the capability of the AI system. 
Our findings suggest that apart from improving user experiences with conversational interfaces, addressing the illusion brought about by such persuasive technologies can be pivotal in facilitating appropriate reliance on AI systems. 
{Systematic empirical explorations are fundamentally important to understand how conversational interfaces can be leveraged effectively to foster optimal human-AI collaboration. In the absence of such efforts, designers and practitioners are often left to make less-informed choices that can lead to unintended consequences. In this spirit,} our work {has important theoretical implications for promoting appropriate reliance using XAI methods, and in equal part, design implications for effective conversational interfaces to support human-AI collaboration. 

\section{Related Work}
This paper focuses on exploring the impact of an XAI dashboard and a conversational XAI interface on user understanding of an AI system (\textbf{RQ1}), which may further affect user trust and appropriate reliance (\textbf{RQ2}).
Thus, we position our work in the following realms of related literature: human-AI decision making (\S\ref{sec-rel-decision-making}),  explainable AI (\S\ref{sec-rel-xai}), and conversational user interfaces (\S\ref{sec-rel-conv-interface}). 

\subsection{Human-AI Decision Making}
\label{sec-rel-decision-making}

While predictive AI systems are powerful, they are seldom perfect~\cite{kocielnik2019will}. 
Transparency and accountability issues prevent deep learning-based AI systems from automation in high-stakes applications like medical diagnosis~\cite{davenport2019potential}. 
In comparison, human workers (\eg medical doctors) show strong reliability and accountability for their work outcomes and decisions, which serve as the foundation for customers to trust their services. 
With these concerns, human-AI collaborative decision making is regarded as a promising approach to taking advantage of both humans and AI to achieve more accurate and reliable decision outcomes.


Complementary team performance is an important goal for human-AI decision making~\cite{bansal2021does,erlei2024understanding}, and will continue to be vital in the age of LLMs~\cite{sau_chi2025,ggu_chi2025,ab_chi2025}. 
To achieve complementary team performance, users of AI systems are expected to rely on AI advice appropriately~\cite{schemmer2022should}. 
To this end, users are expected to follow AI advice when the AI system is more capable than them, and not rely on AI advice when the AI system is less capable. 
When users fail to calibrate their trust in the AI system, they may misuse or disuse the AI advice, resulting in over-reliance and under-reliance, respectively.
The causes for unexpected reliance behaviors are complex. 
For example, algorithm aversion~\cite{dietvorst2015algorithm,erlei2022s}  and algorithm appreciation~\cite{you2022algorithmic} can cause under-reliance and over-reliance, respectively. 
Existing work has extensively explored how confidence~\cite{Zhang-FAT-2020,chong2022human}, risk perception~\cite{green2019disparate,green2020algorithmic}, performance feedback~\cite{Lu-CHI-2021,Rechkemmer-CHI-2022}, and explanations~\cite{wang2021explanations,robbemond2022understanding,erlei2020impact} can affect human-AI decision making. 

Prior studies found that human factors like expertise and domain knowledge~\cite{nourani2020role,Chiang-IUI-2022} and cognitive bias~\cite{bertrand2022cognitive,he2023knowing} can greatly affect user trust~\cite{vereschak2021evaluate} and appropriate reliance~\cite{schemmer2022should} on the AI system. 
To mitigate the negative impact of some human factors, researchers have proposed tutorial interventions~\cite{cai2019hello,Lai-CHI-2020,Chiang-IUI-2022,he2023knowing}, cognitive forcing functions~\cite{buccinca2021trust,lu2024does,he2024err}, and improving transparency of the AI system~\cite{lai2019human,wang2021explanations,Lu-CHI-2021}. \citet{Chiang-IUI-2022} found that a tutorial intervention to reveal the limitations of the AI system can effectively reduce over-reliance. 
Others have explored the role of task factors such as task complexity and uncertainty in shaping trust and reliance in human-AI decision-making~\cite{salimzadeh2024dealing,salimzadeh2024doubt}. \citet{buccinca2021trust} proposed cognitive forcing functions to compel people to engage more thoughtfully with explanations along with AI advice. 
They found that such interventions can effectively mitigate over-reliance. 

In previous work, researchers~\cite{wang2021explanations,warren2023categorical,bhattacharya2023directive,cau2023supporting} explored how different XAI methods may affect user understanding of an AI system, trust, and reliance. 
It is still unclear how the interaction interfaces to present XAI methods will substantially affect user understanding of an AI system, trust, and reliance. 
In this work, we propose to fill in such research gap and explore whether conversational XAI interface can facilitate user understanding of the AI system, which further contributes to increased trust and appropriate reliance.

\subsection{Explainable AI}
\label{sec-rel-xai}
While deep learning-based AI systems have been recognized as powerful predictive toolkits, explainability has been a primary concern that prevents them from becoming widespread practice. 
According to GDPR, users of AI systems have the right to obtain meaningful explanations along with AI predictions~\cite{selbst2018meaningful}. 
Under such circumstances, researchers have proposed a diverse set of XAI methods like feature attribution explanations~\cite{ribeiro2016should,lundberg2017unified}, counterfactual explanations~\cite{wu2021polyjuice}, and contrastive explanations~\cite{jacovi2021contrastive,yin2022interpreting}. 
For a more comprehensive review of existing XAI methods and criteria to evaluate XAI methods, we encourage readers to refer to recent work by~\citet{arrieta2020explainable,nauta2023anecdotal}.


As humans have diverse information needs, there is no one-size-fits-all solution~\cite{liao2021human}. 
With a proposal of putting users/humans at the center of technology design~\cite{wang2019designing,ehsan2020human}, more and more researchers have started to explore human-centered XAI~\cite{liao2021human,ehsan2022human}. 
{In such line of literature, researchers focus on the function of explanation --- how explanations affect user understanding and what characteristics make explanations effective~\cite{yang2022psychological,abdul2020cogam}. 
The mental model~\cite{johnson1980mental} denotes how one person build an internal representation of the external reality,\footnote{\url{https://en.wikipedia.org/wiki/Mental_model}} and plays an important role for analyzing human-centered XAI~\cite{kulesza2012tell,kulesza2013too,bansal2019beyond, rong2023towards}. Through empirical user studies, researchers found that many properties of explanations like {simplicity}~\cite{abdul2020cogam}, {completeness}~\cite{kulesza2013too} will substantially affect user mental model and the effectiveness of explanations.
}


According to Jacovi \etal~\cite{jacovi2023diagnosing}, effective explanations should produce \textbf{coherent} mental models (\ie communicate information which generalizes to contrast cases), be \textbf{complete} to avoid misunderstanding and be \textbf{interactive} to address contradictions. 
We recognize that conversational XAI interfaces can satisfy all the above key properties for providing effective explanations. 
Thus, we argue that a conversational XAI interface may benefit users with a better understanding of the AI system, which can further facilitate user trust and appropriate reliance. 
Existing work has explored conversational XAI interfaces in the contexts of collaborative scientific writing~\cite{shen2023convxai} and decision support with a focus on team performance~\cite{slack2022talktomodel}.  
None of the existing works, however, have systematically explored the impact of conversational XAI interfaces on trust and appropriate reliance. 
To fill this knowledge and empirical gap while complementing existing efforts, we designed a controlled study with loan approval tasks to analyze the impact of a conversational XAI interface on human-AI decision making. 



\subsection{Conversational User Interfaces}
\label{sec-rel-conv-interface}
A conversational user interface (CUI) is a user interface for computers that emulates a conversation with a real human~\cite{wiki:Conversational_user_interface}. 
CUIs have been studied widely across multiple disciplines, such as natural language processing, human-computer interaction, and artificial intelligence. 
Since the famous \textit{Turing Test}~\cite{turing2009computing}, the capability to conduct human-like conversation has for long been recognized as an important property of artificial intelligence. 
Researchers have shown great enthusiasm for developing intelligent conversational user interfaces. 
CUIs have been widely adopted in crowdsourcing~\cite{qiu2020improving}, dialogue systems~\cite{mctear2002spoken}, search engines~\cite{radlinski2017theoretical}, and recommender systems~\cite{jannach2021survey,zhou2021crslab}. 
Nowadays, conversational assistants like Apple Siri, Amazon Alexa, and ChatGPT have shown promising potential in assisting users in their daily life and work. 

The main benefits of conversational user interfaces are the natural interaction experience that they facilitate~\cite{narayanan2002creating}, improved user engagement~\cite{qiu2020improving}, better understandability~\cite{mindlin2024measuring} and accessibility.
Compared with traditional graphical user interfaces (GUIs), CUIs have the advantages of more human-like interaction~\cite{moore2017conversational}, simplifying complex tasks with filtered information~\cite{brennan1990conversation}, and leading to a higher subjective trust in the system~\cite{gupta2022trust}. 
Informed by these prior works, we infer that a conversational XAI interface can have similar advantages over a conventional XAI Dashboard (\ie a GUI to access current XAI methods).
With conversational XAI interfaces, users may better understand the AI system and develop higher trust and more appropriate reliance on the AI system.




Compared with these studies, our focus is to analyze the impact of the XAI interfaces (\ie an XAI dashboard and a conversational XAI interface) on human-AI decision making. 
While several works~\cite{slack2022talktomodel,shen2023convxai,lakkaraju2022rethinking,srivastava2023role} have positioned the conversational XAI interface as a promising direction to support human-AI collaboration, this is still an under-explored research topic that requires more empirical studies. 


\section{Task, Method, and Hypotheses}
In this section, we describe the loan approval task 
and present our hypotheses, which have been preregistered before data collection.

\begin{figure}[htbp]
    \centering
    \includegraphics[width=0.48\textwidth]{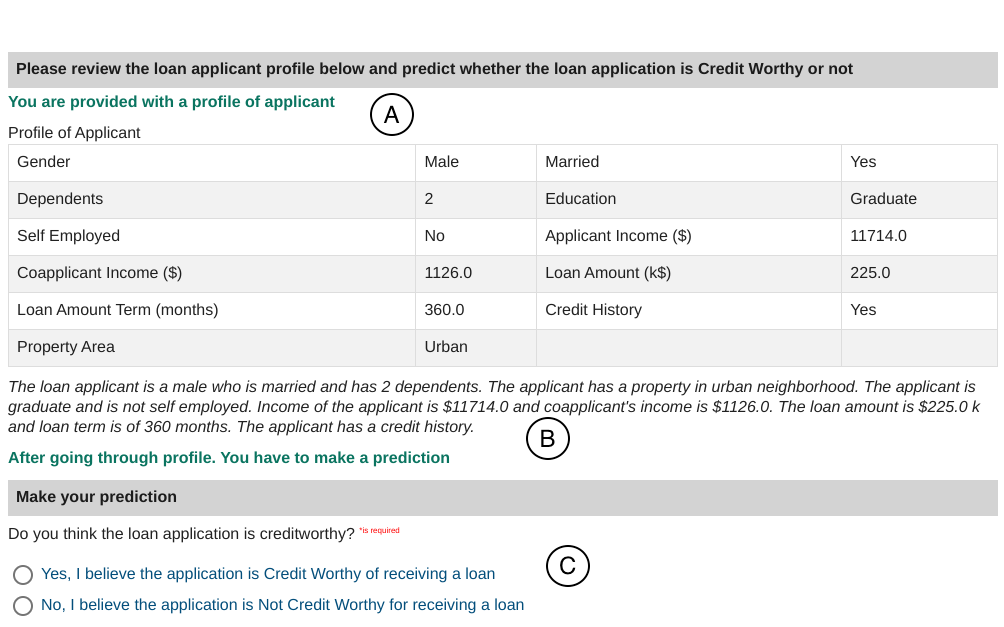}
    \caption{Screenshot of the loan approval task interface. This is the first stage of decision making. (A) Loan Applicant profile is shown in the table with 11 features. (B) To help understand the tabular data, we also provided a textual description below. (C) After going through the profile, participants are asked to decide whether this loan application is `\textbf{Credit Worthy}' or `\textbf{Not Credit Worthy}.'}
    \label{fig:task-interface}
    \Description{A screenshot of task interface. The task interface comprises eleven features (such as income) in a table format. Below, they need to decide whether it is credit worthy or not credit worthy.}
\end{figure}

\begin{table*}[htbp]
	\centering
	\caption{Conversation setup to trigger different XAI responses. Different XAI methods can correspond to different information needs identified in the XAI question bank~\cite{liao2020questioning}. Queries correspond to the options provided in the conversational XAI  interface.}
\label{tab:xai_conversation}%
	\scalebox{.85}{
    \begin{tabular}{p{0.06\textwidth}|p{0.17\textwidth}|p{0.34\textwidth}|p{0.15\textwidth}|p{0.32\textwidth}}
  \toprule
        \textbf{XAI method}& \textbf{Information needs}& \textbf{Queries}&\textbf{User Input}&\textbf{XAI Response}\\
		\hline \hline
		PDP & How& How does [a given feature] influence credit worthiness in general?& Feature Dropdown Selection& Figures illustrating probability distribution when varying specific features and description messages\\
		\hline
		SHAP & Why &What are the most important features influencing the current prediction?& N/A& Figures illustrating the relative importance of all the features and description messages\\
        \hline
        MACE& Why, Why not, How to be
that& What is the minimum change in the applicant's profile needed to switch the current prediction?& N/A& Text Description of minimum change in the profile\\
\hline
		WhatIf& What if, How to be that, How
to still be this& What would happen to the credit worthiness for [a different input]?& Feature Values& Model prediction on a new profile\\
		\hline
        Decision Tree& Why, How to still be this& Which sequence of steps led to the current prediction?& N/A& Figures illustrating the decision path and description message\\
  \bottomrule
    \end{tabular}
    }
\end{table*}%

\subsection{Loan Approval Task}
\label{sec:task}
The basis for our experimental setup is a task where participants have to decide whether a loan application is \textbf{Credit Worthy} or \textbf{Not Credit Worthy} using the publicly available loan prediction dataset.\footnote{\url{https://www.kaggle.com/altruistdelhite04/loan-prediction-problem-dataset}} 
The rationale for selecting the loan approval task as a test bed is three-fold. 
Firstly, this task was chosen as a critical decision making scenario for human-AI collaboration, where there is a clear risk and a benefit when adopting AI advice. 
Secondly, most laypeople are familiar with this context and can make informed decisions based on their knowledge. 
Thirdly, It has also been adopted by existing research in behavioral economics~\cite{bertrand2006behavioral} and human-AI collaboration~\cite{green2019principles,he2023stated}.

In the loan approval task, participants are presented with eleven features (including loan amount, income, and the absence or presence of credit history) in both table format and text description (as shown in Figure~\ref{fig:task-interface}). 
Based on the application profile (composed of the eleven features), participants are asked to decide whether the loan applicant is credit worthy to get the loan approved.
This simulates a realistic scenario where participants interact with an AI system and may rely on AI advice and XAI methods due to the inherent complexity in decision-making~\cite{salimzadeh2023missing}. 
{As the selected loan approval task is one where decision making is fully based on the eleven features, it would be easier to assess users' decision criteria based on the top-ranked features explicitly specified by the users themselves.}

\paratitle{Two-stage Decision Making}. In our study, we adopted a two-stage decision making process for each loan approval task. 
Every participant in our study is first asked to work on the loan approval task without any assistance from the AI system. 
After that, they were given a second chance to alter their initial choice according to the AI advice (\ie AI prediction) and AI explanations (\eg XAI dashboard, according to different experimental conditions). 
This setup is similar to the update condition in work by \citet{green2019principles}. 
{This setup is apt for analyzing user incorporation of system advice and user trust in the AI system~\cite{green2019disparate,dietvorst2018overcoming}. 
It is a widely adopted setup in empirical studies exploring human-AI decision making~\cite{wang2021explanations,Lu-CHI-2021,Chiang-IUI-2022,he2023knowing}.
To assess user decision criteria, we ask users to indicate the three most important features influencing their decision at each stage 
along with their confidence in each decision.}

\subsection{Design of XAI Interfaces}
\label{sec:cxai}


\paratitle{XAI methods}. {Our selection of XAI methods is informed by the taxonomy of XAI methods regarding user information needs~\cite{nauta2023anecdotal,wang2021explanations,liao2020questioning}. Following the XAI question bank~\cite{liao2020questioning}, we selected six user information needs associated with the rationale of AI advice: \textit{how} (global model-wide explanation), \textit{why, why not, how to be that} (a different prediction), \textit{how to still be this} (current prediction), and \textit{what if}. These user information needs can be addressed with five widely-used XAI methods (correspondence summarized in Table~\ref{tab:xai_conversation}).}
{These are (1) A global explanation method -- PDP (\ie partial dependency plot)~\cite{friedman2001greedy}, which visualizes how one feature globally impacts the model prediction, (2) Feature importance attribution method -- SHAP~\cite{lundberg2017unified}. Based on Shapley values, the SHAP method provides feature importance to indicate how each feature supports or opposes the current model prediction. 
(3) Counterfactual explanation method -- MACE~\cite{yang2022mace}. 
MACE will inform users of the minimum changes in the applicant profile required to flip model prediction. 
(4) Widely adopted interactive XAI toolkit -- WhatIf.\footnote{\url{https://pair-code.github.io/what-if-tool/}} Based on the WhatIf toolkit, users can modify the applicant profile and obtain the model prediction for the new profile. (5) Decision tree-based explanation.\footnote{\url{https://scikit-learn.org/stable/modules/generated/sklearn.tree.DecisionTreeClassifier.html}} 
This is one popular XAI method, which makes decisions based on a tree-structure decision criteria. 
In our implementation, we provide the decision path to reach the AI advice.} 
We implemented all these XAI methods by using the OmniXAI library.\footnote{\url{https://github.com/salesforce/OmniXAI}} {More details can be found in supplementary materials.}

\begin{figure*}[htbp]
 \centering
  \subfigure[XAI Dashboard with WhatIf Response.]{\label{fig:dashboard}
  \centering
  \includegraphics[width=0.48\textwidth]{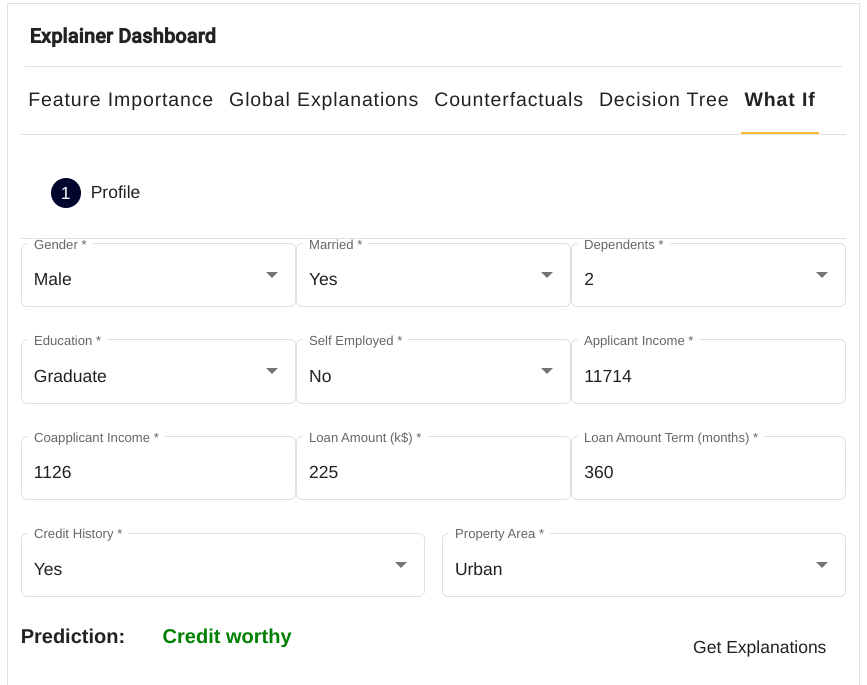}
 }
  \subfigure[Conversational XAI interface with SHAP Response.]{\label{fig:personalization-shap}
  \centering
  \includegraphics[width=0.48\textwidth]{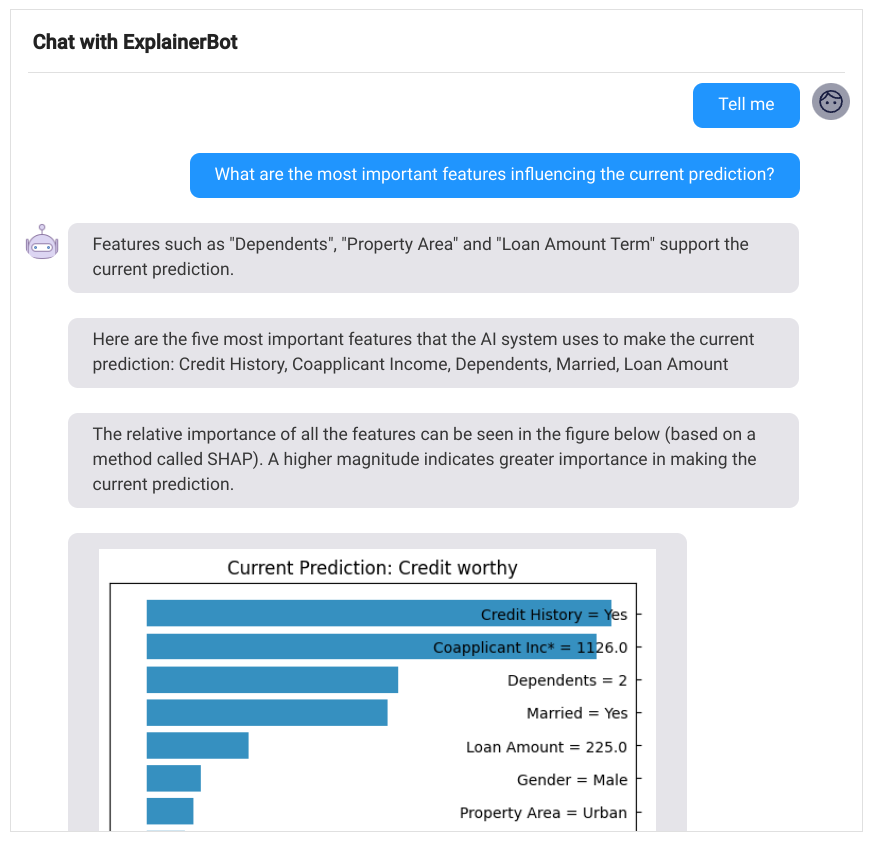}
 }
 \centering
 \caption{Screenshots illustrating the XAI interfaces we designed. Additional screenshots demonstrating all XAI methods across both XAI interfaces are available in the supplementary materials.}
 \label{fig-xai-interface}
 \Description{Two screenshots of XAI interfaces used in our study. (a) XAI dashboard with WhatIf explainer and (b) a personalized conversational interface with SHAP explainer.}
\end{figure*}

\paratitle{XAI Dashboard}. 
{Following existing standards, the XAI dashboard is an interactive interface that provides users with XAI responses on demand when accessed through the navigation tab (see Figure~\ref{fig:dashboard}). Users can explore all XAI methods by focusing on one at a time, which ensures both simplicity and complete coverage of the available five XAI methods.}

\paratitle{Conversational XAI Interface}. 
Templating conversational interactions via a rule-based agent~\cite{grudin2019chatbots} can be an effective method to guide users in exploring their information needs and understanding the model decisions. 
Thus, we adopted a rule-based conversational agent to power the conversational XAI interface.
By referring to the XAI question bank~\cite{liao2020questioning}, we first set up five user intents (see Table~\ref{tab:xai_conversation}), which can be answered with the corresponding XAI responses.

To provide a smooth conversational experience, we curated the five user intents into three categories: about AI advice (SHAP, MACE, Decision Tree --- XAI responses required no user input), AI advice for modified applicant profile (WhatIf, where users need to revise the applicant profile), and the global impact of a specific feature (PDP, where users need to specify a feature of interest). 
At the beginning of the conversation, users are guided to select one category among the three and then specify one query to check or specify user input. After users receive one XAI response, we repeat the aforementioned process.
All user intents are wrapped into an iterative loop, and users can stop the conversation after receiving at least two different XAI responses. 
All the conversations are guided by empowering participants to select options using custom buttons and commands (\ie dropdown selection for PDP or feature input for WhatIf, shown in Figure~\ref{fig:dashboard}). {Such designs have been widely adopted in domains such as conversational crowdsourcing~\cite{qiu2020improving,qiu2020ticktalkturk}, or customer service chatbots and proven to be effective in addressing user information needs and are easy to use for laypeople~\cite{klopfenstein2017rise}.} 

\paratitle{Evaluative Conversational XAI Interface for Decision Support}. 
Based on the collected user decision criteria in the initial decision, we further adapted the conversation to guide users to check such features (\ie top-$3$ features selected in the initial decision making). 
{This is inspired by the evaluative AI for explainable decision support~\cite{miller2023explainable}, which argues for `providing evidence for and against decisions made by people.' 
Such evaluative conversational XAI interfaces nudge users to think about their initial decision criteria further by comparing them with explanations from AI systems. 
To this end, it is similar to cognitive forcing functions~\cite{buccinca2021trust}, which has been adopted to calibrate user trust and reliance behaviors.}

To achieve the goal of {evaluative decision support} in our conversational XAI interface, we adopted guiding messages in the customized buttons with \revise{user decision criteria} ( \ie the top-$3$ features the user selected in initial decision making). 
For XAI methods that require user input (\ie PDP and WhatIf), we adapted the guiding message with \revise{user decision criteria}. 
For example, instead of selecting one option for PDP, users have an extra option to directly explore how one of the selected features influences credit worthiness. 
We believe that doing so can help them to explore how the selected features will affect the model prediction. 
After obtaining the XAI response, the conversational assistant sends a message to check whether the user wants to continue exploring the current XAI method by either modifying or selecting a feature randomly sampled from \revise{user decision criteria}. 
In the case of SHAP, MACE, and Decision Tree {(\ie XAI methods which do not require user input)}, the conversational assistant sends a message about how their {initial decision criteria} 
work in current XAI methods{, serving as evaluative feedback}. 
Similarly, this message helps them to check how their decision criteria differ from the AI system (as reflected by explanations provided via the XAI methods). 
After users obtain {the SHAP, MACE, or Decision Tree} XAI response, the conversational assistant provides {an extra option} message to guide them to explore the PDP (\ie global explanation on feature variation) response with one randomly selected feature from their initial set of top-$3$ features. 




\paratitle{\revise{Conversational XAI Interface with LLM Agents}}.\footnote{\revise{To notice that, the conversational XAI interface supported with LLM agents was adopted as a follow-up comparison with other conditions. In the pre-registration, we only include samples and hypotheses associated with other XAI interfaces.}} 
\revise{
While rule-based agents can inform the flow in conversational interactions, 
they lack the flexibility to deal with user needs in a bilateral human-like conversation.
To address such concerns and further our understanding of the impact of flexible interaction and enhanced conversation quality in the conversational XAI interface, we built another conversational XAI interface powered by LLM agents. 
The benefits of introducing LLM agents are two-fold: (1) LLMs have shown promising user query understanding capability, which enables understanding user information needs and generating coherent and high-quality personalized conversation responses~\cite{zhao2023survey}. (2) When equipped with XAI methods as potential tools, LLM agents can provide suitable XAI responses on demand, which may provide a better user experience (\eg more flexible expression of information needs and high-quality text responses based on XAI outcomes). 
}

\revise{Apart from the difference in agents (LLM agents in this case), the entire procedure is identical to the basic conversational XAI interface. 
Our implementation of the LLM agent is based on autogen~\cite{wu2023autogen} and GPT-4. 
Given user queries, the LLM agent-based conversational XAI  transforms user intents into pre-defined explainers and elaborates on the generated explanations to generate coherent text responses. 
We also provide the five hint questions (as shown in Table~\ref{tab:xai_conversation}) to trigger potential XAI responses during the conversation in a randomized order on every task. 
Users can ask the LLM agent any questions using textual input.  
For more implementation details of our LLM agent-based conversational XAI interface, readers can refer to our supplementary materials. 
}


\subsection{Hypotheses}
\label{sec:hypo}


Our experiment was designed to answer questions surrounding the impact of 
conversational XAI interfaces on user understanding, trust, and reliance on AI systems. 
{XAI dashboards, which can switch between different XAI methods with a navigation bar, have been recognized as a promising interactive interface to present explanations towards model decisions~\cite{oege_dijk_2022_6408776,spinner2019explainer,slack2023explaining}. 
Considering its wide application for model explainability, we consider it a strong baseline in our study.}
As shown in prior work, conversational user interfaces have the advantages of more human-like interaction~\cite{moore2017conversational} and simplified understanding of complex tasks with filtered information~\cite{brennan1990conversation} over graphical user interfaces. 
Compared with the XAI dashboard (where users interact with the dashboard in a uni-lateral fashion), the conversational XAI interface has the potential to increase user engagement, and provides a more natural bi-directional way for users to explore their information needs and develop an understanding of the AI system. 
As a result, users with a conversational XAI interface may develop a better understanding of the AI system. 
Thus, we hypothesize that:

\begin{framed}
\noindent\textbf{(H1)}: Compared to the XAI dashboard, the conversational XAI interface creates a better understanding of the AI system among users.
\end{framed}


Prior work has highlighted that humans show higher trust when interacting with intelligent systems using a conversational interface compared to conventional web interfaces~\cite{gupta2022trust}. 
Further, conversational user interfaces have been shown to increase worker engagement in microtask crowdsourcing~\cite{qiu2020improving} compared to a traditional GUI. 
Such increased engagement can potentially help users deliberate, reflect, and thereby make better decisions, relying on the AI system more critically. 
{Conversational XAI interfaces can help users explore and address different information needs, which may bring a higher trust in the AI system.}
Thus, we hypothesize:

\begin{framed}
\noindent\textbf{(H2)}: Compared to the XAI dashboard, the conversational XAI interface will help users exhibit a relatively higher trust in the underlying AI system. \\
\textbf{(H3)}: Compared to the XAI dashboard, the conversational XAI interface will help users exhibit a relatively more appropriate reliance on the underlying AI system.
\end{framed}

{Evaluative decision support} in the XAI interface may further help users reassess their initial thoughts about the AI system and AI advice. 
By revealing the difference among their decision criteria and providing explanations for the AI system's advice, users can obtain a better understanding of the AI system and make more critical decisions~\cite{miller2023explainable}. 
This can in turn facilitate critical thinking about the AI system, leading to a potential calibration of user trust and increased appropriate reliance on the AI system. 
Thus, we hypothesize that:

\begin{framed}
\noindent\textbf{(H4)}: Adaptive steering of conversations {for evaluative decision support} in the conversational XAI interface will increase user trust and appropriate reliance on an AI system.
\end{framed}
\section{Study Design}
This section describes our experimental conditions, variables, and procedures related to our study. This study was approved by the human research ethics committee of our institution.

\subsection{Experimental Conditions}
The main aspects of our research questions and hypotheses concern the effect of different XAI interfaces. 
In our study, all participants worked on the loan approval tasks with a two-stage setup (described in Section~\ref{sec:task}), where AI advice is provided in the second stage of decision making. 
The only difference is the nature of the interface through which AI advice is explained. 
Considering this factor as the sole independent variable in our study, we designed a between-subjects study with \revise{five} experimental conditions:
\begin{itemize}
    \item \control{}: no XAI interface.
    \item \dashboard{}: with XAI dashboard interface (as described in Section~\ref{sec:cxai}).
    \item \chatxai{}: with a conversational XAI interface (as described in Section~\ref{sec:cxai}).
    \item \chatpersonalized{}: with a {evaluative} conversational XAI interface (as described in Section~\ref{sec:cxai}).
    \item \revise{\chatagent{}: with a conversational XAI interface powered by LLM agents (as described in Section~\ref{sec:cxai}).}
\end{itemize}


\subsection{Measures and Variables}
Our hypotheses mainly considered five types of dependent variables: user understanding, user trust, performance, reliance, and appropriate reliance on the AI system.

\paratitle{User Understanding of the AI System}. 
{This work focuses on analyzing the impact of the XAI interfaces instead of evaluating the quality of explanations~\cite{holzinger2020measuring}. 
In our study, user understanding of the AI system is a function of interactive exploration with the XAI interfaces, which can evolve while working on tasks. Note that we consider and describe perceived explanation utility as a separate construct below.} 
Based on existing literature~\cite{shin2021effects,szymanski2021visual,bove2022contextualization,schmude2023impact}, we synthesized and adopted {four} dimensions to assess user understanding of the AI system. 
As a result of practice through our study, users can potentially learn across tasks and understand the system. We aim to capture this through the dimensions of \textit{Perceived Feature Understanding}, \textit{Learning Effect} across tasks, and \textit{Understanding of the System}. 
All questionnaires used to assess user understanding can be found in supplementary materials.
To objectively quantify user understanding of the features, we calculated nDCG~\cite{jarvelin2017ir} of users' top-$3$ features and the SHAP feature importance ranking as \textit{Objective Feature Understanding}. 
For the relevance scores, we adopted a decreasing relevance for the SHAP feature order (based on the abstract value of SHAP values) with an interval of 1. 
Thus the relevance scores range from [1, 11] for the 11 features we used. 
Besides, \textit{Perceived Feature Understanding} is also used as an indicator of perceived user understanding.

\paratitle{Explanation Utility}. {Alongside user understanding, the perceived explanation utility is an important aspect identified in the existing literature on human-centered XAI~\cite{rong2023towards,ehsan2022human,ehsan2021operationalizing,liao2021human}. 
We synthesized and adopted four dimensions based on existing literature to evaluate the explanation utility provided in conditions with XAI interface. 
According to Jacovi \etal~\cite{jacovi2023diagnosing}, effective explanations can provide users with a coherent and complete mental model to explain the current AI prediction. 
Thus, we adopted the dimensions of \textit{Explanation Completeness} and \textit{Explanation Coherence} in our post-task questionnaires. 
According to \citet{hsiao2021roadmap}, perceived \textit{Explanation Clarity} and \textit{Explanation Usefulness} are also important dimensions for assessing perceived explanation goodness.}

\paratitle{User Trust}. 
\citet{mohseni2021multidisciplinary} showed that understandability and predictability are desired properties for trustworthy intelligent systems. 
Moreover, the perceived competence of the AI system (\ie users' confidence about the system's capabilities) 
and reliability of the AI system (\ie the extent to which the system is perceived not suffer from unexpected errors) are also identified as essential constructs to establish trust~\cite{ryan2020ai,twyman2008trust}. 
In addition to capturing these attributes, we also captured subjective trust of users by adopting 
three validated subscales from the trust in automation questionnaire~\cite{korber2019theoretical}.
These are TiA-Reliability/Competence (TiA-R/C), TiA-Understanding/Predictability (TiA-U/P), and TiA-Trust in Automation (TiA-Trust). 
Each subscale is calculated as the average score (5-point Likert) across related questions. 
These measures have been shown to be meaningful to use in empirical studies of human-AI decision making~\cite{lai2021towards,he2023stated}.

\paratitle{Performance and Reliance}. 
As has been argued by prior work, assessing user reliance on the AI system when users agree with AI advice can be inaccurate~\cite{schemmer2022should}. 
Thus, we measure both performance and user reliance from two distinct standpoints. 
Besides the global user performance (\ie overall \textit{Accuracy}), we also considered user performance when their initial choice disagreed with AI advice (\ie \textit{Accuracy-wid}). 
Similarly, we consider \textit{Agreement Fraction} (\ie how often users agree with AI advice in their final decisions) as a global measure of reliance. We consider \textit{Switch Fraction} (\ie how often users adopt AI advice in cases of initial disagreement) as another precise indicator of user reliance.
To assess appropriate reliance, we followed Max \etal~\cite{schemmer2022should} to adopt Relative positive AI reliance (\textit{RAIR}) and Relative positive self-reliance (\textit{RSR}) metrics. 
These measures enumerate all cases when the user initially disagrees with AI advice, but the correct decision is present in one of them. 
By calculating the positive reliance patterns among all potential actions, \textit{RAIR} and \textit{RSR} assess whether users know when they should rely on the AI system and themselves, respectively. 
To our knowledge, they are the most representative objective measures of appropriate reliance.

\paratitle{Other Variables}. To dive deep into the impact of different XAI interfaces, we also considered other variables in our study. 
User confidence has been identified as an important factor in human-AI decision making~\cite{passi2022overreliance,chong2022human,green2019principles}. 
In our study, we recorded user confidence in each stage of decision making tasks with the question--``\textit{What is your confidence level while making this decision?}.'' 
As described in Section~\ref{sec:hypo}, the conversational XAI interface may benefit human-AI decision making with higher user engagement. 
To quantitatively analyze such impact, we adopted the UES-SF~\cite{o2018practical} questionnaire in our study and considered the average score across all dimensions as an indicator of user engagement.  

\subsection{Procedure}
\begin{figure}[htbp]
    \centering
    \includegraphics[width=0.48\textwidth]{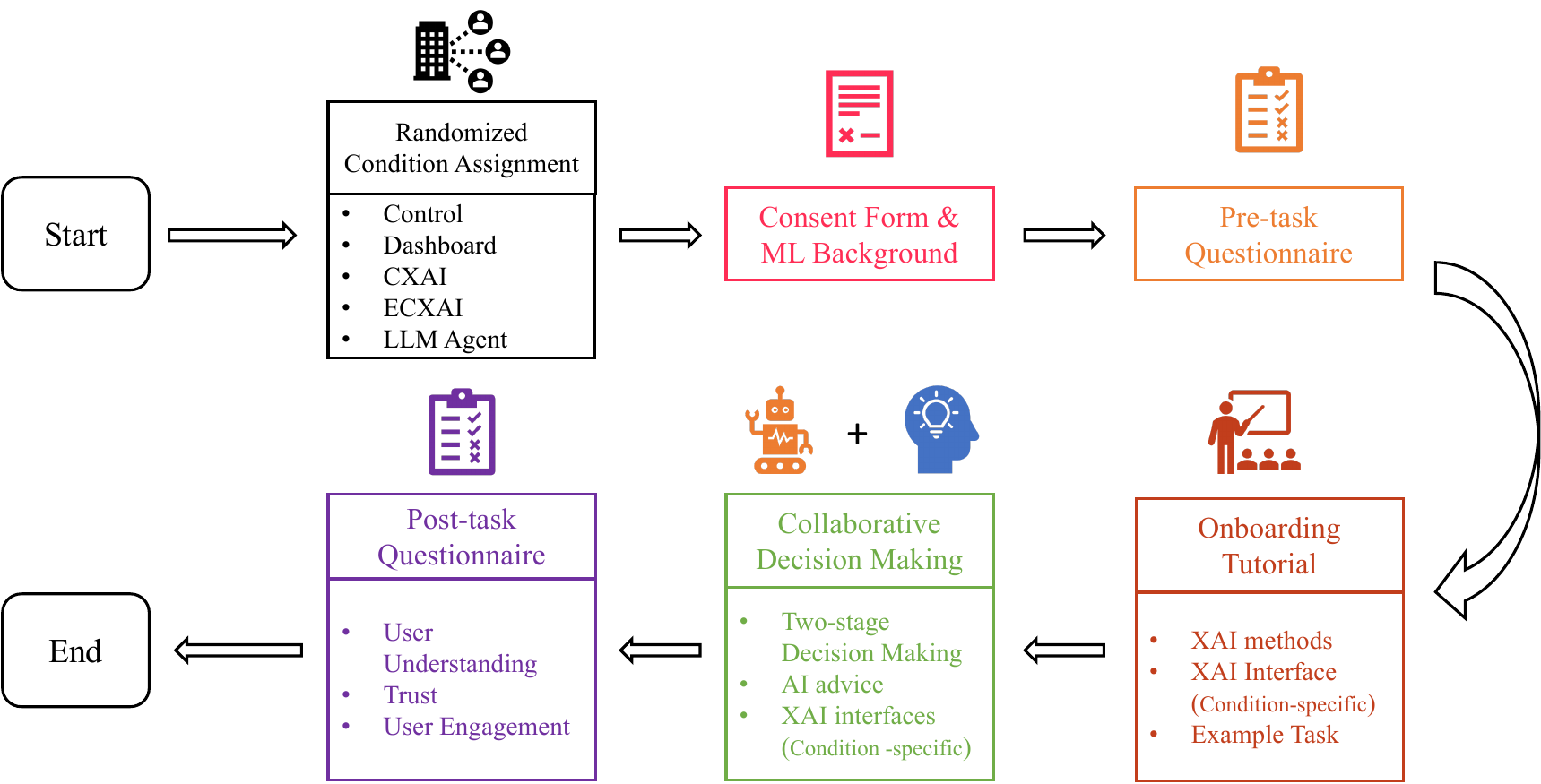}
    \caption{Illustration of the procedure that participants followed in our study. This flow chart describes the experimental condition \chatxai{}.}
    \label{fig-procedure}
    \Description{A flow chart to illustrate the procedure participants experienced in our study. It consists of six steps: (1) randomized condition assignment, (2) consent form and machine learning background, (3) pre-task questionnaire for ATI, (4) onboarding tutorial, (5) collaborative decision making with two-stage setup, and (6) post-task questionnaire.}
\end{figure}
The complete procedure participants followed in our study is illustrated in Figure~\ref{fig-procedure}. All participants will be first randomly assigned to one experimental condition. 
To proceed with participation, all participants were first asked to sign an informed consent form by clicking a button and also indicate their prior experience with machine learning. 
Next, participants were asked to complete a pre-task questionnaire to measure their affinity for technology interaction (\ie ATI).
%
Then, an onboarding tutorial and a practice example were provided to help participants get familiar with the two-stage decision making setup and the corresponding XAI interface depending on the experimental condition.\footnote{More details pertaining to the onboarding tutorial can be found in the supplementary material.}
At this stage, participants in the  \control{} condition only see one practice example to get familiar with the loan approval task. 
Participants then worked on the ten selected tasks within a two-stage decision making setup. 
Finally, they were asked to fill in post-task questionnaires (including the TiA questionnaire and questions pertaining to user understanding of the AI system via the XAI methods).
\section{Experimental Results}
In this section, we present the results of our empirical study. \revise{In addition to the main results, we carried out exploratory analyses to draw nuanced interpretations of our key insights.
Readers can refer to the appendix.} 
Our code and data can be found at Github.\footnote{\url{https://github.com/delftcrowd/IUI2025_ConvXAI}}

\subsection{Descriptive Statistics}
To ensure the reliability of our results and interpretations, we only consider participants who passed all attention checks. 
Finally, the participants considered for analysis were distributed in a balanced manner across the four experimental conditions: 61 (\control{}), 61 (\dashboard{}), 62 (\chatxai{}), 61 (\chatpersonalized{}), 
 \revise{61 (\chatagent{}). On average, each task consumes 13 API calls to obtain responses in \chatagent{} condition, including generating reply messages and XAI usage. The average time (mins) spent across conditions are: 22 (\control{}), 34 (\dashboard{}), 52 (\chatxai{}), 45 (\chatpersonalized{}), 62 (\chatagent{}). With Kruskal-Wallis H-tests and post-hoc Mann–Whitney test, we confirmed significance: \control{} < \dashboard{} < \chatxai{}, \chatpersonalized{} < \chatagent{}.} 

\paratitle{Distribution of Covariates}. The covariates’ distribution is as follows: 
\textit{ML Background} (\revise{$22.5\%$} with machine learning background knowledge, \revise{$77.5\%$} without machine learning background knowledge), 
\textit{ATI} (\revise{$M = 3.99$, $SD = 0.90$}; 6-point Likert scale, \textit{1: low, 6: high}), 
\textit{TiA-Propensity to Trust} (\revise{$M = 2.88$, $SD = 0.71$}; 5-point Likert scale, \textit{1: tend to distrust}, \textit{5: tend to trust}), 
and \textit{TiA-Familiarity} (\revise{$M = 2.67$, $SD = 1.10$}; 5-point Likert scale, \textit{1: unfamiliar}, \textit{5: very familiar}). 

\paratitle{Performance Overview}. On average across all conditions, participants achieved an accuracy of
\revise{$64.5\%$ ($SD = 0.11$)}, which is still lower than the AI accuracy ($70\%$). The agreement fraction is \revise{0.847 ($SD = 0.16$)}, and the
switching fraction is \revise{0.522 ($SD = 0.41$)}. With these measures, we confirm that when users disagree with AI advice, they do not always blindly rely on AI advice. 
As all dependent variables are not normally distributed, we used non-parametric statistical tests to verify our hypotheses.

\begin{figure}[htbp]
    \centering
    \includegraphics[width=0.49\textwidth]{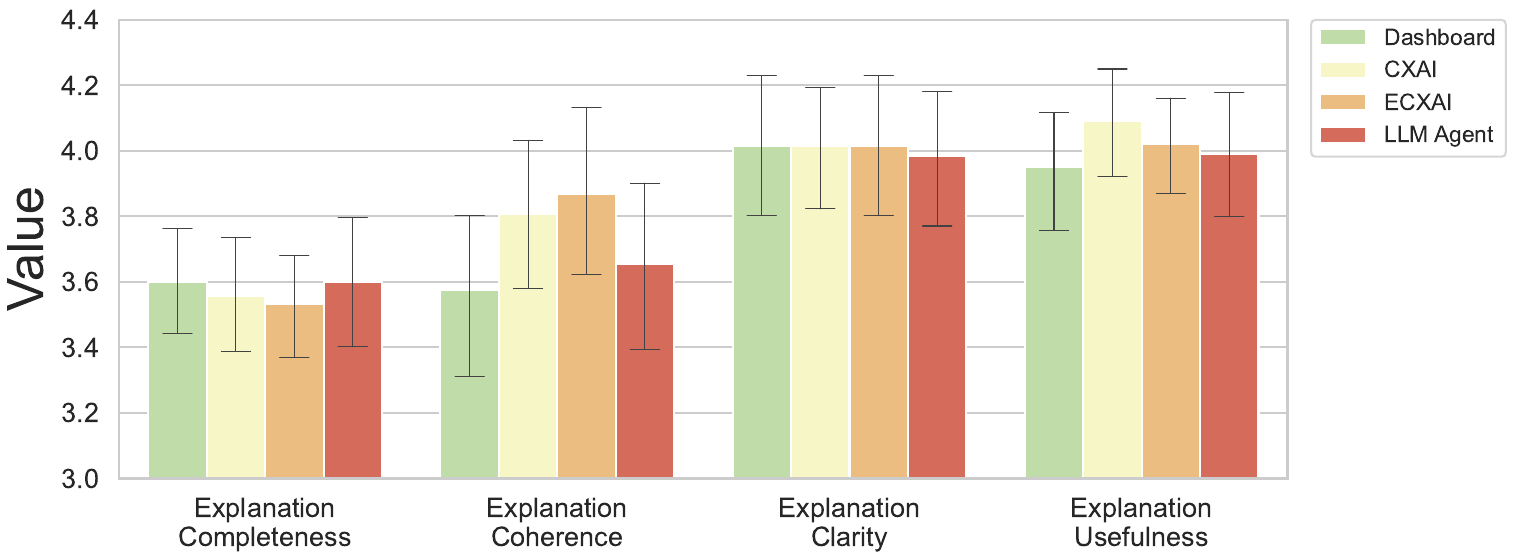}
    \caption{Bar plot illustrating the explanation utility across conditions. Error bars represent the $95\%$ confidence interval.}
    \label{fig:explanation_utility}
    \Description{Bar plots for explanation utility (Explanation Completeness, Explanation Coherence, Explanation Clarity, and Explanation Usefulness) across conditions with XAI interfaces. Overall, all conditions achieved similar Explanation Completeness and Explanation Clarity. Meanwhile, the conversational XAI interfaces achieved slightly better Explanation Coherence and Explanation Usefulness.}
\end{figure}

\paratitle{Explanation Utility}. To illustrate how the XAI interface will affect the perceived explanation utility, we adopted a bar plot of explanation utility across conditions. As shown in Figure~\ref{fig:explanation_utility}, participants achieved similar level of \textit{Explanation Completeness} and \textit{Explanation Clarity}. Meanwhile, participants with conversational XAI interfaces (\ie condition \chatxai{}, \chatpersonalized{}, and \chatagent{}) achieved slightly higher \textit{Explanation Coherence} and \textit{Explanation Usefulness}. Based
on one-way ANOVA, we analyzed the impact of XAI interfaces in perceived explanation utility. There is no significant difference across conditions.




\subsection{Hypothesis Tests}
\label{sec-hyp-test}
For the convenience of the readers, we have provided concise insights in the main body of this section and placed additional tables and figures (\eg estimation plots) that provide further details in the supplementary materials. 


\begin{table*}[hbpt]
	\centering
	\caption{Kruskal-Wallis H-test results for XAI interfaces (\textbf{H3} and \textbf{H4}) on reliance-based dependent variables. {The post-hoc results are based on Mann–Whitney tests.} ``${\dagger\dagger}$'' indicates the effect of the variable is significant at the level of 0.0125.
 }
	\label{tab:res-reliance}%
	\begin{small}
        \scalebox{0.9}{
        \revise{
	\begin{tabular}{c | c c | c c c c c| c}
	    \hline
	    \multirow{2}{*}{\textbf{Dependent Variables}}& \multirow{2}{*}{$H$}& \multirow{2}{*}{$p$}& \multicolumn{5}{c|}{$M \pm SD$}& \multirow{2}{*}{Post-hoc results}\\
     \cline{4-8}
     &  & & \control& \dashboard{}& \chatxai{}& \chatpersonalized{}& \chatagent{}& \\
	    \hline \hline
	    \textbf{Accuracy}& 9.09 & .059& $0.62 \pm 0.13$ &$0.65 \pm 0.11$ &$0.67 \pm 0.10$ &$0.64 \pm 0.09$ &$0.63 \pm 0.10$& -\\
        \hline
	\textbf{Agreement Fraction}& 33.66 & \textbf{.000}$^{\dagger\dagger}$& $0.74 \pm 0.17$ &$0.86 \pm 0.17$ &$0.89 \pm 0.15$ &$0.85 \pm 0.16$ &$0.89 \pm 0.11$ & \control{} < \dashboard{}, \chatxai{}, \chatpersonalized{}, \chatagent{} \\\hline
	\textbf{Switch Fraction}& 19.14 & \textbf{.001}$^{\dagger\dagger}$& $0.31 \pm 0.34$ &$0.57 \pm 0.41$ &$0.58 \pm 0.43$ &$0.57 \pm 0.41$ &$0.57 \pm 0.41$ & \control{} < \dashboard{}, \chatxai{}, \chatpersonalized{}, \chatagent{}\\\hline
    \textbf{Accuracy-wid}& 5.06 & .281& $0.46 \pm 0.30$ &$0.50 \pm 0.36$ &$0.52 \pm 0.35$ &$0.55 \pm 0.38$ &$0.42 \pm 0.36$ & -\\\hline
    \textbf{RAIR}& 11.01 & \textbf{.026}$^{\dagger}$& $0.35 \pm 0.39$ &$0.50 \pm 0.44$ &$0.60 \pm 0.45$ &$0.52 \pm 0.44$ &$0.48 \pm 0.45$ & \control{} < \chatxai{} \\\hline
	\multirow{2}{*}{\textbf{RSR}}& \multirow{2}{*}{38.26} & \multirow{2}{*}{\textbf{.000}$^{\dagger\dagger}$}& \multirow{2}{*}{$0.57 \pm 0.46$} &\multirow{2}{*}{$0.29 \pm 0.44$} & \multirow{2}{*}{$0.23 \pm 0.40$} & \multirow{2}{*}{$0.26 \pm 0.41$} & \multirow{2}{*}{$0.11 \pm 0.29$} &  \control{} > \dashboard{}, \chatxai{}, \chatpersonalized{}, \chatagent{}\\
    & & & & & & & & \dashboard{} > \chatagent{}\\
	    \hline
	\end{tabular}
    }
    }
	\end{small}
\end{table*}

\subsubsection{\textbf{H1}: effect of XAI interfaces on user understanding}
\label{sec-h1}
To analyze the main effect of the XAI interfaces on user understanding of the AI system, we conducted an \textit{Analysis of Covariance} (ANCOVA) with the \textit{experimental condition} as between-subjects factor and \textit{TiA-Propensity to Trust}, \textit{TiA-Familiarity}, \textit{ATI}, and \textit{ML Background} as covariates. 
While our data may not be normally distributed, we still adopted  AN(C)OVAs for analysis because these analyses have been shown to be robust to Likert-type ordinal data~\cite{Norman-2010-likert}. 
For this analysis, we considered all participants across three experimental conditions with XAI (\ie \dashboard{}, \chatxai{}, and \chatpersonalized{}).  
We found 
no significant differences resulting from the 
different XAI interfaces (\ie experimental condition). However, the \textit{TiA-Propensity to Trust} showed a significant impact on all dimensions of user understanding. 
For the objective feature understanding (continuous value, non-normal distribution), we conducted Kruskal-Wallis H-tests by considering different XAI interfaces. \revise{A significant difference ($H = 16.19, p = .001$) was found between participants with different XAI interfaces. Through post-hoc Mann–Whitney U test, we found that \chatagent{} condition achieved significantly worse \textit{objective feature understanding} than the \dashboard{}, \chatxai{}, and \chatpersonalized{} conditions.}
Thus, we did not find any support for \textbf{H1}.

\subsubsection{\textbf{H2}: effect of XAI interfaces on user trust}
\label{sec-h2}

To verify \textbf{H2} (\ie the impact of XAI interface on user trust), we conducted an 
\textit{Analysis of Covariance} (ANCOVA) with the \textit{experimental condition} as between-subjects factor and \textit{TiA-Propensity to Trust}, \textit{TiA-Familiarity}, \textit{ATI}, and \textit{ML Background} as covariates. 
This allows us to explore the main effects of the XAI interface on subjective trust as measured by the three subscales of the Trust in Automation questionnaire~\cite{korber2019theoretical}. 

As we found, the experimental condition (\ie XAI interface) only showed a significant impact in \textbf{TiA-U/P}. 
With post-hoc Tukey's HSD test, we found that participants who received XAI showed significantly higher trust in \textbf{Understandability/Predictability} (\ie \control{} < \dashboard{}, \chatxai{}, \chatpersonalized{}). 
\revise{Besides the significant results, participants in the \chatagent{} condition showed a consistent but non-significant trend across all measures: \control{} < \chatagent{} < \dashboard{}, \chatxai{}, \chatpersonalized{}.} 
However, no significant difference is found between the \dashboard{} condition and conditions with conversational XAI. 
At the same time, there is no significant impact of the experimental conditions observed on the dependent variables of \textbf{TiA-R/C} and \textbf{TiA-Trust}. 
Meanwhile, we found that \textbf{TiA-Propensity to Trust} had a significant impact on all trust-related dependent variables, and that users' affinity to technology interaction (\textbf{ATI}) also had a significant impact on \textbf{TiA-U/P}.

To better understand effect sizes in terms of the TiA-U/P and go beyond \textit{p}-values, we adopted an estimation plot~\cite{ho2019moving} (shown in supplementary materials, Figure 3). As reflected by the swarm plot, participants with conversational XAI interface (\ie condition \chatxai{} and \chatpersonalized{}) exhibited a marginally higher TiA-U/P in comparison with condition \dashboard{}. 
Thus, we found partial support for \textbf{H2}.  


\subsubsection{\textbf{H3}: effect of XAI interfaces on appropriate reliance} 
\label{sec-h3}
To verify \textbf{H3}, we conducted a Kruskal-Wallis H-test to compare the performance, reliance, and appropriate reliance measures of participants across four experimental conditions. 
As shown in Table~\ref{tab:res-reliance}, participants showed significantly higher reliance (\ie \textit{Agreement Fraction} and \textit{Switch Fraction}) with access to the XAI dashboard or conversational XAI interface. 
However, the increased reliance is not necessarily appropriate reliance. 
Only participants with access to conversational XAI interface (\ie condition \chatxai{}) showed significantly better \textit{RAIR} in comparison with the condition \control{}.
We also found that participants showed significantly worse \textit{RSR} with access to the XAI dashboard or conversational XAI interface. 
\revise{We also notice that participants in the \chatagent{} condition showed significantly worse RSR compared to the \control{} and \dashboard{} conditions, which indicates that the \chatagent{} condition led to severe over-reliance on the AI advice.} 
Thus, \textbf{H3} is not supported by our experimental results.

There is no significant difference in team performance (\ie \textit{Accuracy} and \textit{Accuracy-wid}). 
To interpret our data beyond \textit{p}-values and better understand effect sizes in terms of the overall team performance,
we adopted estimation plots~\cite{ho2019moving}  {(shown in supplementary materials, Figure 4)}. Based on the normal distribution sampled for these measures, we can infer the reliance difference based on the mean difference of the estimated distribution.
We found that: (1) Compared to the \control{} condition, participants 
in the \chatxai{} condition 
showed a clearly higher mean accuracy.
(2) Participants in the \chatpersonalized{} condition showed slightly better \textit{Accuracy-wid} than the \dashboard{} condition and the \chatxai{} condition.
Similarly, we adopted estimation plots~\cite{ho2019moving} {(cf. supplementary materials, Figure 4)} to draw meaningful interpretations related to our appropriate reliance measures. 
We found that: (1) Compared to the \control{} condition, 
participants in the \chatxai{} condition showed a significantly higher  \textit{RAIR}.
At the same time, participants in the \chatxai{} condition showed a slightly higher \textit{RAIR} compared with participants in  \dashboard{} and  \chatpersonalized{} conditions.
(2) Participants in the \dashboard{} and \chatpersonalized{} conditions showed slightly better \textit{RSR} than the \chatxai{} condition.

\subsubsection{\textbf{H4}: effect of {evaluative conversation} on user trust and appropriate reliance}

According to results reported in Section~\ref{sec-h2} and Section~\ref{sec-h3}, 
no significant difference in user trust and appropriate reliance was found between experimental condition \chatxai{} and \chatpersonalized{}. Thus, \textbf{H4} is not supported. 
\section{Discussion}

\subsection{Key Findings}
Our experimental results show that participants with an interactive XAI interface ({\ie either an XAI dashboard or a conversational XAI interface}) can obtain a relatively high degree of perceived understanding, trust, and reliance on the AI system. 
However, the increase in trust and reliance may potentially stem from an illusion of their understanding of explanatory depth~\cite{chromik2021think,rozenblit2002misunderstood}. 
As a result, they do not necessarily know when the AI advice is trustworthy and worth relying on. 
This is reflected by the over-reliance we observed (see Table~\ref{tab:res-reliance}) in all conditions with interactive XAI interfaces. 
with an LLM agent-based conversational XAI interface (Section~\ref{sec-h3}), we observed that over-reliance was further reinforced (\ie worse \textit{RSR}) and users obtained significantly worse \textit{objective feature understanding} compared to other conditions with XAI interfaces. 
This indicates that instead of calibrating user trust and reliance on the AI system, {enhancing the conversation quality} may further induce the illusion of explanatory depth.

\paratitle{Positioning in Existing Literature}. In our study, we found that interactive XAI interfaces 
can have a negative impact of increasing over-reliance on the AI system. 
This is consistent with the findings of previous empirical studies of human-AI collaboration~\cite{Zhang-FAT-2020,lai2021towards,wang2021explanations}. 
Our results indicate that participants perceive the conversational XAI interface to lead to a relatively better user understanding and team performance than the XAI dashboard. 
This is in line with findings of Slack \etal~\cite{slack2023explaining}, where they found TalktoModel (a conversational XAI interface) was preferred by most participants and achieved better team performance when collaborating with users. 
We extend existing empirical work by going one step further to explore the impact of conversational XAI interfaces on trust and appropriate reliance. 
We found that users tend to show relatively higher trust and appropriate reliance on the conversational XAI interface.
Further {enhancement} of the conversation {(\ie adaptive steering for evaluative decision support)} 
does not necessarily help further improve user understanding, user trust, and appropriate reliance on the AI system (\ie the \chatpersonalized{} and \chatagent{} conditions). 
Instead, we found that it can even be harmful (cf. Section~\ref{sec-hyp-test}), which is reflected by a decreased user understanding of the AI system, user trust, and appropriate reliance in the \chatagent{} condition.
Our exploratory findings 
suggest promising avenues for future research --- further exploring how conversational XAI interfaces can affect user trust and reliance on the AI system through additional confirmatory studies in different contexts. 
Our work is an important first exploration to this end, and 
more empirical studies are required to corroborate and further contextualize these observations. As we strive towards optimal human-AI decision making, we highlight an important trade-off that needs to be managed between creating {user-friendly, seamless, and plausible} conversational XAI interfaces and simultaneously fostering critical consideration of AI advice.

\subsection{Implications of Our Work}


\paratitle{{Interactive XAI Interfaces Can Amplify Illusions of Explanatory Paths}}. 
{Our work has important theoretical implications for promoting appropriate reliance on AI systems with XAI methods.} 
In our study, participants with the XAI dashboard as well as the conversational XAI interfaces showed obvious over-reliance on the AI system. 
The reason behind this can be that participants with XAI interfaces developed illusions of the intelligence level of the AI system. \revise{Prior work has shown that conversational interfaces can build user trust~\cite{gupta2022trust}, and XAI can bring about an illusion of explanatory depth~\cite{chromik2021think}. Both can contribute to uncalibrated trust in the AI system and cause over-reliance. Their combination could potentially amplify users’ over-reliance depending on other task, human, and system factors.} 
As our results suggested, participants with conversational XAI interface (\ie \chatxai{}) showed slightly better perceived user understanding across multiple dimensions (non-significant results) and trust (\ie \revise{Understanding/Predictability}) than participants with XAI dashboard. 
At the same time, participants in condition \chatxai{} also showed the best \textit{RAIR} and relatively worse \textit{RSR} 
(see Table~\ref{tab:res-reliance}), while participants in the \chatagent{} condition showed the worst \textit{RSR} (see Section~\ref{sec-h3}). 
{Combined with exploratory findings in Table~\ref{tab:exploratory_reliance_correlation} --- user understanding{, explanation utility,} and user trust is positively correlated with over-reliance.} 
This indicates that the conversational XAI interface appears to be more persuasive to users and leads to relatively more over-reliance on the AI system. {Thus, optimizing the XAI interfaces as a persuasive technology~\cite{fogg2002persuasive} may not be the ideal approach to promoting appropriate reliance on AI systems. In extreme cases, persuasive technology can even help untrustworthy AI systems deceive end users to gain their trust~\cite{banovic2023being}. 
Instead, we should focus on developing methods and interfaces that can ensure that the XAI responses provided will not mislead users by creating an illusion of system intelligence or explanatory depth.} 

\paratitle{Towards more effective conversational XAI interfaces}. 
{Our work has important implications for designing effective conversational XAI interfaces.} 
{Rather than being persuasive, we expect effective XAI interfaces to be accessible and low-barrier interfaces that can enhance user engagement and guide users to explore their information and explanation needs. 
As a result, users can have a better user experience, and a more comprehensive understanding of the AI system (\eg including both strengths and weaknesses), resulting in more appropriate reliance on the AI system. 
}
In our study, the conversational XAI interface failed to facilitate a significantly better user understanding, trust, and appropriate reliance. 
Based on our findings, there are multiple potential approaches to improve the effectiveness of the conversational XAI interface. 

Firstly, the trustworthiness of AI advice should be calibrated within the conversation. 
As we found, the improved user experience and conversation quality do not necessarily translate into appropriate reliance. 
To that end, users 
need to be supported with faithful conversations, which may help them realize whether AI advice is trustworthy. 
To tackle the vulnerability of improved plausibility (\eg introducing LLMs or other persuasive technology), future work can explore how to align the trustworthiness of AI advice with the plausibility of conversational XAI responses. 
\revise{Secondly, conversational XAI interfaces could be used to address potential issues associated with AI literacy. 
Conversational interactions have been proven to be effective in supporting novice and low-literacy users in using mobile interfaces~\cite{medhi2011designing}. Prior work has shown that AI literacy plays an important role in calibrating user trust and reliance behavior~\cite{Chiang-IUI-2022}. Thus, leveraging conversational XAI interfaces to narrow down the literacy gap when working with AI systems can also be a promising future direction to explore.}
{\revise{Thridly}, although adaptive evaluative steering for evaluative decision support fails to facilitate optimal human-AI decision making, it leads to substantial impacts on user perception and user reliance behavior. 
For example, participants in condition \chatpersonalized{} achieved slightly higher \textit{Explanation Coherence}, slightly higher \textit{Accuracy-wid} and decreased \textit{Agreement Fraction} compared to condition \chatxai{}. 
Such an evaluative AI~\cite{miller2023explainable} conceptual framework could still be a promising approach to facilitating human-AI interaction within a conversational manner. Future work can further combine such evaluative conversational XAI with cognitive forcing functions~\cite{buccinca2021trust} through the dialogue to help calibrate user trust and reliance.
} 
{Similarly, Ehsan \etal~\cite{ehsan2022seamful} proposed the framework of Seamful XAI to augment explainability and user agency in human-AI collaboration by revealing the ``seams'' (\ie imperfections of the AI system). 
Combined with these ideas, we can guide users to explore both the strengths and weaknesses of the AI system.} 
Such a conversation may be more engaging and may potentially achieve similar functions as cognitive forcing functions~\cite{buccinca2021trust} to help participants make decisions more critically. This is an important direction for future work.

\subsection{Caveats and Limitations}

In our study, we selected the most representative five XAI methods as the basis to form our interactive XAI interfaces. 
We cannot overrule that this design choice may have been a bottleneck for some participants in our study, as they may have had information needs that are not covered by the XAI methods. Once users find that their queries cannot be answered properly based on pre-defined XAI methods, their trust and reliance on the AI system may decrease.
Having said that, our setup is representative of current state-of-the-art AI-assisted decision making methods. 
In our study, the conversational XAI interfaces in the \chatxai{} and \chatpersonalized{} conditions are built upon rule-based dialogue systems. 
All conversations are guided in a pre-defined manner, which lacks flexibility in communication. 
We 
developed an LLM agent-based conversational XAI interface (\ie the \chatagent{} condition) to select XAI methods on demand, improve the scope and quality of user interactions, and flexibly communicate the corresponding explanations.
We found that {more flexible and plausible} conversations did not necessarily help further improve user trust and appropriate reliance on the AI system. 
Instead, it amplified over-reliance and negatively impacted user understanding of the AI system.
Based on these results, we can infer that, improving the conversational quality by using more human-like utterances may be more persuasive and strengthen the illusion of explanatory depth. 

According to prior studies about crowdsourcing~\cite{gadiraju2015understanding}, some participants can rush through the study and provide low-effort results. 
To alleviate participants with low-effort results, we adopted attention checks in the questionnaire and tasks in our study.
Meanwhile, it would be challenging to keep participants engaged in the XAI interface and highly motivated to learn from the explanations of XAI responses. 
To ensure that participants spent enough effort to interact with the conversational XAI interface, participants were required to view at least two different types of XAI responses in each conversation. This was, however, not explicitly mentioned and participants were alerted to this only when they tried to proceed without engaging with the XAI methods. 

\paratitle{{Broader Societal Implications}}. 
Our findings add to the urgency to be careful when employing AI-based decision support systems due to their tendency to act as persuasive technologies. 
Although evaluative conversations led to an increase in user trust and reliance in our study, contrary to expectations, this did not amount to an increased appropriate reliance. Future work can explore similar `evaluative AI'~\cite{miller2023explainable} operationalizations in conversational human-AI interaction and decision support. 
We found that users' propensity to trust is strongly correlated with their subjective trust in the AI system and their appropriate reliance (cf. Section~\ref{sec-h1} and covariate analysis in supplementary materials). 
Participants with a higher propensity to trust showed significantly higher trust and reliance (\ie \textit{Agreement Fraction} and \textit{Switch Fraction}) on the AI system. 
As a result, they were more likely to develop an illusion of explanatory depth and over-rely on misleading AI advice. 
Such a tendency to trust may have originated from a lack of AI literacy~\cite{Chiang-IUI-2022} and a critical mindset~\cite{he2024err}. These results, along with recent findings in the IUI community~\cite{Chiang-IUI-2022} suggest that the development and deployment of AI systems and XAI interfaces can systematically favor individuals with higher AI literacy or critical mindsets, and therefore cause disparities to others. Further work is required to ensure that different types of users (with varying AI literacy or differing individual traits) can equally benefit from AI systems and related interfaces. 
\section{Conclusion}
In this paper, we presented a first-of-its-kind empirical study to understand the impact of an XAI dashboard and a conversation XAI interface on user understanding of the AI system, and their further impact on user trust and appropriate reliance. 
Compared to participants with the XAI dashboard, participants with the conversational XAI interface showed a slightly better understanding (\textbf{RQ1}), and demonstrated a slightly higher trust in the AI system (\textbf{RQ2}). 
However, our findings suggest that the XAI interfaces were persuasive and have the potential to bring about an illusion of the AI systems' capability, which in turn increased over-reliance on the AI system. 
Moreover, we found that {evaluative} conversational interactions do not work as expected in facilitating user trust and understanding. 
With \revise{experimental results associated with} conversational XAI interfaces powered with LLM agents, we found that {boosting} the conversation {quality and flexibility (\ie with LLM-based conversational agent)} may further reinforce over-reliance and hurt user understanding and user trust.
Our insights and observations can inform the future design of conversational XAI interfaces to promote complementary human-AI collaboration. Conversational XAI interfaces should balance user engagement with seamful design requirements that can promote decision making that is married with critical reflection.

Our results indicate that we should be careful in presenting XAI methods with an interactive XAI interface, which may cause over-reliance on the AI system. 
While our experimental results do not provide support to our original hypotheses, more work is required to further contextualize the effectiveness of conversational XAI interfaces in shaping user understanding, trust, and appropriate reliance. 
As opposed to further improving user experiences with conversational XAI interfaces in the context of human-AI decision making, future work should first focus on mitigating the illusion of explanatory depth brought by the XAI methods.

\begin{acks}
This work was partially supported by the Delft Design@Scale AI Lab, the 4TU.CEE UNCAGE project, and the Convergence Flagship ``ProtectMe'' project. We made use of the Dutch national e-infrastructure with the support of the SURF Cooperative using grant no. EINF-5571 and EINF-9738. We finally thank all participants from Prolific and experts from our department.
\end{acks}

\bibliographystyle{ACM-Reference-Format}
\bibliography{convxai}

\clearpage
\appendix

\section{Appendix}

\subsection{Implementation Details}

\paratitle{Task Selection}. All participants in our study were presented with ten loan approval tasks in the main task phase. 
All such cases are selected from the test set of a random split of the full dataset (training / test ratio 4:1). 
All tasks were evenly split between those where the loan applicant should be \textbf{Credit Worthy} (\textbf{CW}) for the loan being approved and those where the applicant profile should be \textbf{Not Credit Worthy} (\textbf{NCW}). 
As shown in Table~\ref{tab:selected_tasks}, we selected the ten tasks according to prediction correctness and model confidence. 
We first trained an XGBoost Classifier~\cite{chen2016xgboost} based on the training set. 
For both \textbf{CW} cases and \textbf{NCW} cases, we selected one high-confidence correct prediction, one random-confidence correct prediction, one low-confidence correct prediction, and one high-confidence wrong prediction. While we adopted another random-confidence correct prediction for class \textbf{NCW}, we selected another low-confidence wrong prediction for class \textbf{CW} to control the accuracy of the AI system to be $70\%$. 
This {experimental} design was also informed by a pilot study without AI advice. 
We recruited 20 participants from the Prolific platform to work on the selected loan approval tasks, and found that they achieved an accuracy level around $60\%$. 
To ensure the AI system is helpful to improve human decision making accuracy and maintain the risk of accepting wrong advice, we manually controlled the accuracy of the AI system to be $70\%$. 
During the study, we randomly shuffled the task order for each participant to prevent ordering effects \cite{nourani2021anchoring}. 

\begin{table}[htbp]
	\centering
	\caption{Task selection criteria for our study. `CW' and `NCW' refer to \textbf{Credit Worthy} and \textbf{Not Credit Worthy}, respectively.}
	\label{tab:selected_tasks}%
	\begin{small}
	\begin{tabular}{c | c | c | c }
	    \toprule
	    \textbf{Task ID}& \textbf{Groud Truth}& \textbf{Correctness}& \textbf{Model Confidence}\\
	    \hline \hline
            1& CW& $\checkmark$& High\\
            2& CW& $\checkmark$& Low\\
            3& CW& $\checkmark$& Random\\
            4& CW& $\times$& Low\\
            5& CW& $\times$& High\\
            \midrule
            6& NCW& $\checkmark$& High\\
            7& NCW& $\checkmark$& Low\\
            8& NCW& $\checkmark$& Random\\
            9& NCW& $\checkmark$& Random\\
            10& NCW& $\times$& High\\
	    \bottomrule
	\end{tabular}%
	\end{small}
\end{table}%

\paratitle{Sample Size Estimation}. 
To ensure that our empirical study has a sufficient sample size for statistical analysis, we computed the required sample
size in a power analysis for a Between-Subjects ANOVA using G*Power~\cite{faul2009statistical}. 
To correct for testing multiple hypotheses, we applied a Bonferroni correction so that the significance threshold decreased to $\frac{0.05}{4}=0.0125$. 
We specified the default effect size $f = 0.25$, a significance threshold $\alpha = 0.0125$ (\ie due to testing multiple hypotheses), a statistical power of $(1 - \beta) = 0.8$, and that we will investigate four different experimental conditions/groups. This resulted in a required sample size of $244$ participants. 
We thereby recruited participants from the crowdsourcing platform Prolific.\footnote{\url{https://www.prolific.co}} As illustrated in Figure \ref{fig-procedure}, participants were recruited continuously and randomly assigned to an experimental condition, simultaneously accommodating for potential exclusion until the required sample size was reached (as described below). As a result, 352 participants were recruited \revise{for conditions \control{}, \dashboard{}, \chatxai{}, and \chatpersonalized{}}, of which 107 were excluded. \revise{In the experiment process, the \chatagent{} condition was considered as a follow-up study, which is not included in the initial sample size estimation. For ease of comparison with other conditions, we recruited 61 valid participants for \chatagent{} condition.}

\paratitle{Compensation}. All participants were rewarded with \pounds 4, amounting to an hourly wage of \pounds 8 deemed to be a ``\textit{good}'' payment by the platform (estimated completion time was 30 minutes). 
On top of this basic payment, we rewarded participants with extra bonuses of \pounds 0.05 for every correct decision in the ten loan approval tasks. 
This bonus setting encourages participants to reach a correct decision to the best of their ability, which is also a contextual requirement to encourage appropriate system reliance~\cite{lee2004trust}.

\paratitle{Filter Criteria}. All participants were proficient English speakers above the age of 18, and had
finished over 40 tasks while maintaing an approval rate of over 90\% on the Prolific platform. 
To ensure reliable participation, we employed attention check questions (one for decision making, three for questionnaires) in our study. 
All attention check questions explicitly direct participants to select a specific option. 
They were designed to look similar to the questions or decision making tasks they were embedded in~\cite{gadiraju2015understanding}. 
If users read our instructions and engaged genuinely with the task, passing these attention check questions is straightforward. 
We excluded participants from our analysis if they failed at least one attention check or if we found any missing data. 
The resulting sample of \revise{306} participants had an average age of \revise{32} (SD = 7.8) and a
gender distribution (\revise{53.6\% female, 46.4\% male}).

\paratitle{Questionnaire}. To assess the user understanding of the AI system and explanation utility, we collected questionnaires shown below from participants:

\begin{small}
\begin{itemize}[leftmargin=*, nosep]
\item \textbf{Perceived Feature Understanding}: \\
\textit{1. The explanations helped you improve and/or reinforce your understanding of the influential features.}\\$\Box$ Strongly disagree $\Box$ Disagree $\Box$
  Neutral $\Box$ Agree $\Box$ Strongly Agree

\item \textbf{Understanding of the System}\\
\textit{1. I can understand why the system provided specific explanations.}\\$\Box$ Strongly disagree $\Box$ Disagree
  $\Box$ Neutral $\Box$ Agree $\Box$ Strongly Agree

\item \textbf{Learning Effect across Tasks}\\
\textit{1. My understanding of AI system and decision criteria improve over the tasks.}\\$\Box$ Strongly disagree $\Box$ Disagree
  $\Box$ Neutral $\Box$ Agree $\Box$ Strongly Agree

\end{itemize}
\end{small}
To assess the explanation utility, we collected questionnaires shown below from participants:

\begin{small}
\begin{itemize}[leftmargin=*, nosep]
\item \textbf{Explanation Completeness}\\
\textit{1. The explanations provide a sufficient rationale that supports the AI advice.
}\\$\Box$ Strongly disagree $\Box$ Disagree
  $\Box$ Neutral $\Box$ Agree $\Box$ Strongly Agree\\
  \textit{2. The explanations sufficiently express the uncertainty of the AI advice.
}\\$\Box$ Strongly disagree $\Box$ Disagree
  $\Box$ Neutral $\Box$ Agree $\Box$ Strongly Agree

\item \textbf{Explanation Coherence}\\
\textit{1. The explanations you received are consistent with your initial expectations.
}\\$\Box$ Strongly disagree $\Box$ Disagree
  $\Box$ Neutral $\Box$ Agree $\Box$ Strongly Agree

\item \textbf{Explanation Usefulness}\\
\textit{1. The provided explanations are useful in making final decision.}\\$\Box$ Strongly disagree $\Box$ Disagree
  $\Box$ Neutral $\Box$ Agree $\Box$ Strongly Agree

\item \textbf{Explanation Clarity}\\
\textit{1. Explanations are clear enough to inform my final decision.}\\$\Box$ Strongly disagree $\Box$ Disagree
  $\Box$ Neutral $\Box$ Agree $\Box$ Strongly Agree

\end{itemize}
\end{small}

\subsection{Additional Exploratory Analyses}

\subsubsection{Impact of Covariates}
As shown in the analysis for \textbf{H2} (cf. Table~\ref{tab:correlation}), covariates like TiA-Propensity to Trust and ATI have shown some impact on user trust. 
To further analyze the impact of covariates on human-AI decision making, we conducted Spearman rank-order tests between covariates and all categories of dependent variables. The results are shown in Table~\ref{tab:correlation}. 
We have the following main findings: (1) Overall, \textit{TiA-Propensity to Trust} significantly positively impacted most dependent variables in user understanding, trust, and reliance categories. 
(2) While the propensity to trust positively correlated with user reliance (\ie \textit{Agreement Fraction} and \textit{Switch Fraction}), it negatively affects \textit{RSR}. In other words, some participants with a higher propensity to trust tend to over-rely on the AI system. 
(3) \textit{TiA-Familiarity} and \textit{ATI} only showed some positive impact on user understanding and user trust. No significant correlation was found for user reliance. 
(4) \revise{\textit{ML background} showed positive correlation with user trust. Meanwhile, some dimensions of explanation understanding also show a borderline positive correlation}

\begin{table*}[htbp]
	\centering
	\caption{Correlation of covariates and dependent variables. ``$\dagger$'' and ``$\dagger\dagger$'' indicate the effect of the variable is significant at the level of 0.05 and 0.0125, respectively.
 }
	\label{tab:correlation}%
	\begin{small}
	\begin{tabular}{r | c c | c c| c c | c c }
	    \hline
	    \textbf{Covariates}&	\multicolumn{2}{c|}{\textbf{Propensity to Trust}}& \multicolumn{2}{c|}{\textbf{TiA-Familiarity}} & \multicolumn{2}{c|}{\textbf{ATI}} & \multicolumn{2}{c}{\textbf{ML background}}\\
	    \hline
	    \textbf{Dependent Variables}& $r$& $p$& $r$& $p$& $r$& $p$& $r$& $p$ \\
	    \hline \hline
            Perceived Feature Understanding & 0.344 &\textbf{.000}$^{\dagger\dagger}$& 0.131 & .041$^{\dagger}$& 0.148 & .021$^{\dagger}$& 0.049 & .444\\
            Explanation Completeness & 0.366 &\textbf{.000}$^{\dagger\dagger}$& 0.106 & .097& 0.073 & .254& 0.152 & .017$^{\dagger}$\\
            Explanation Coherence & 0.387 &\textbf{.000}$^{\dagger\dagger}$& 0.131 & .040$^{\dagger}$& 0.087 & .175& 0.135 & .035$^{\dagger}$\\
            Explanation Clarity & 0.427 &\textbf{.000}$^{\dagger\dagger}$& 0.069 & .285& 0.129 & .044$^{\dagger}$& 0.142 & .026$^{\dagger}$\\
            Learning Effect Across Tasks & 0.232 &\textbf{.000}$^{\dagger\dagger}$& 0.173 &\textbf{.007}$^{\dagger\dagger}$& 0.115 & .072& 0.147 & .021$^{\dagger}$\\\
            Understanding of System & 0.343 &\textbf{.000}$^{\dagger\dagger}$& 0.082 & .202& 0.146 & .022$^{\dagger}$& 0.080 & .210\\
            Explanation Usefulness & 0.423 &\textbf{.000}$^{\dagger\dagger}$& 0.166 &\textbf{.009}$^{\dagger\dagger}$& 0.172 &\textbf{.007}$^{\dagger\dagger}$& 0.083 & .196\\
            Objective Feature Understanding & 0.108 & .092& -0.152 & .017$^{\dagger}$& 0.013 & .844& -0.024 & .714\\
            \hline
            TiA-R/C & 0.677 &\textbf{.000}$^{\dagger\dagger}$& 0.126 & .028$^{\dagger}$& 0.171 &\textbf{.003}$^{\dagger\dagger}$& 0.153 &\textbf{.008}$^{\dagger\dagger}$\\
            TiA-U/P & 0.472 &\textbf{.000}$^{\dagger\dagger}$& 0.083 & .150& 0.243 &\textbf{.000}$^{\dagger\dagger}$& 0.158 &\textbf{.006}$^{\dagger\dagger}$\\
            TiA-Trust & 0.774 &\textbf{.000}$^{\dagger\dagger}$& 0.235 &\textbf{.000}$^{\dagger\dagger}$& 0.154 &\textbf{.007}$^{\dagger\dagger}$& 0.164 &\textbf{.004}$^{\dagger\dagger}$\\
            \hline
            Accuracy & 0.091 & .111& 0.073 & .202& -0.039 & .502& -0.019 & .740\\
            Agreement Fraction & 0.223 &\textbf{.000}$^{\dagger\dagger}$& 0.055 & .335& 0.030 & .598& -0.039 & .499\\
            Switch Fraction & 0.137 & .016$^{\dagger}$& -0.030 & .595& -0.001 & .982& 0.037 & .518\\
            Accuracy-wid & 0.056 & .326& 0.032 & .582& -0.045 & .434& 0.057 & .322\\
            RAIR & 0.118 & .040$^{\dagger}$& -0.001 & .980& -0.026 & .648& 0.026 & .654\\
            RSR & -0.186 &\textbf{.001}$^{\dagger\dagger}$& -0.024 & .674& -0.080 & .162& -0.038 & .505\\
	    \hline
	\end{tabular}%
	\end{small}
\end{table*}%

\subsubsection{{The Impact of User Perceptions on Their Behavior}} Prior work has shown that user trust can substantially affect user reliance behaviors~\cite{lee2004trust,Tolmeijer-UMAP-2021}. 
To further analyze how {perception-based variables (\ie user trust, user understanding, and explanation utility)} affect team performance and user reliance behaviors, we conducted Spearman rank-order tests between {corresponding categories of variables}.
The results are presented in Table~\ref{tab:exploratory_reliance_correlation}. 

\begin{table*}[htbp]
	\centering
	\caption{Correlation between {perception-based variables (\ie user understanding, explanation utility, and user trust) and behavior-based variables}. ``$\dagger$'' and ``$\dagger\dagger$'' indicate the effect of the variable is significant at the level of 0.05 and 0.0125, respectively.
 }
	\label{tab:exploratory_reliance_correlation}%
	\begin{small}
	\begin{tabular}{r | c c | c c| c c | c c | c c | c c}
	    \hline
	    \textbf{Behavior-based Variables}&	\multicolumn{2}{c|}{\textbf{Accuracy}}& \multicolumn{2}{c|}{\textbf{Accuracy-wid}} & \multicolumn{2}{c|}{\textbf{Agreement Fraction}} & \multicolumn{2}{c|}{\textbf{Switch Fraction}}& \multicolumn{2}{c|}{\textbf{RAIR}} & \multicolumn{2}{c}{\textbf{RSR}}\\
	    \hline
	    \textbf{Perception-based Variables}& $r$& $p$& $r$& $p$& $r$& $p$& $r$& $p$& $r$& $p$& $r$& $p$ \\
	    \hline \hline
            Perceived Feature Understanding& 0.045 & .484& -0.024 & .709& 0.254 &\textbf{.000}$^{\dagger\dagger}$& 0.117 & .067& 0.096 & .135& -0.293 &\textbf{.000}$^{\dagger\dagger}$\\
            Objective Feature Understanding & 0.332 &\textbf{.000}$^{\dagger\dagger}$& 0.195 &\textbf{.002}$^{\dagger\dagger}$& 0.469 &\textbf{.000}$^{\dagger\dagger}$& 0.322 &\textbf{.000}$^{\dagger\dagger}$& 0.269 &\textbf{.000}$^{\dagger\dagger}$& -0.297 &\textbf{.000}$^{\dagger\dagger}$\\
            Learning Effect Across Tasks & 0.084 & .192& -0.085 & .184& 0.170 &\textbf{.008}$^{\dagger\dagger}$& -0.006 & .931& 0.007 & .913& -0.135 & .035$^{\dagger}$\\
            Understanding of System & 0.114 & .076& -0.083 & .197& 0.157 & .014$^{\dagger}$& 0.010 & .877& -0.017 & .795& -0.153 & .016$^{\dagger}$\\
            \hline
            Explanation Completeness & 0.050 & .435& 0.056 & .387& 0.146 & .022$^{\dagger}$& 0.142 & .026$^{\dagger}$& 0.157 & .014$^{\dagger}$& -0.170 &\textbf{.007}$^{\dagger\dagger}$\\
            Explanation Coherence& 0.107 & .095& -0.030 & .643& 0.270 &\textbf{.000}$^{\dagger\dagger}$& 0.068 & .286& 0.005 & .935& -0.218 &\textbf{.001}$^{\dagger\dagger}$\\
            Explanation Clarity & 0.002 & .973& -0.111 & .083& 0.190 &\textbf{.003}$^{\dagger\dagger}$& 0.081 & .204& 0.042 & .514& -0.235 &\textbf{.000}$^{\dagger\dagger}$\\
            Explanation Usefulness & 0.125 & .051& 0.081 & .206& 0.361 &\textbf{.000}$^{\dagger\dagger}$& 0.266 &\textbf{.000}$^{\dagger\dagger}$& 0.229 &\textbf{.000}$^{\dagger\dagger}$& -0.300 &\textbf{.000}$^{\dagger\dagger}$\\
            \hline
            TiA-R/C& 0.127 & .047$^{\dagger}$& 0.090 & .162& 0.224 &\textbf{.000}$^{\dagger\dagger}$& 0.195 &\textbf{.002}$^{\dagger\dagger}$& 0.175 &\textbf{.006}$^{\dagger\dagger}$& -0.200 &\textbf{.002}$^{\dagger\dagger}$\\
            TiA-U/P & 0.099 & .123& 0.051 & .430& 0.210 &\textbf{.001}$^{\dagger\dagger}$& 0.132 & .038$^{\dagger}$& 0.125 & .051& -0.182 &\textbf{.004}$^{\dagger\dagger}$\\
            TiA-Trust & 0.145 & .024$^{\dagger}$& 0.032 & .617& 0.254 &\textbf{.000}$^{\dagger\dagger}$& 0.164 &\textbf{.010}$^{\dagger\dagger}$& 0.152 & .017$^{\dagger}$& -0.203 &\textbf{.001}$^{\dagger\dagger}$\\
	    \hline
	\end{tabular}%
	\end{small}
\end{table*}%

We found that: (1) \textit{Agreement Fraction} and \textit{RSR} are significantly correlated with most dimensions of user understanding, explanation utility, and user trust. However, these dimensions are positively correlated with \textit{Agreement Fraction} but negatively correlated with \textit{RSR}. 
{This suggests that the improved user understanding, {explanation utility,} and user trust with XAI interfaces can partially explain the increased over-reliance on the AI system.} 
(2) While user trust dimension \textit{TiA-R/C} and \textit{TiA-Trust} positively correlated with reliance measures (\textit{Agreement Fraction} and \textit{Switch Fraction}), and \textit{RAIR}, they negatively correlated with \textit{RSR}. As a result, they do not show a significant correlation with \textit{Accuracy-wid}. 
{This corroborates that higher user trust in the AI system does not necessarily translate into appropriate reliance behaviors.}
(3) Overall, \textit{Objective Feature Understanding} seems useful to facilitate appropriate reliance. 
With a higher \textit{objective Feature Understanding}, participants demonstrate better team performance and higher reliance. 
Although it still contributes to over-reliance (reflected by negative correlation with \textit{RSR}), it shows a more positive impact on appropriate reliance (\ie \textit{Accuracy-wid} and \textit{RAIR}). {In comparison, the positive impact of \textit{Explanation Usefulness}, \textit{TiA-R/C}, and \textit{TiA-Trust} on mitigating under-reliance (\ie positive correlation with \textit{RAIR}) get canceled by the side effect of over-reliance (\ie negative correlation with \textit{RSR}). As a result, these variables do not significantly contribute to team performance.}

\begin{figure*}[h]
    \centering
    \includegraphics[width=0.9\textwidth]{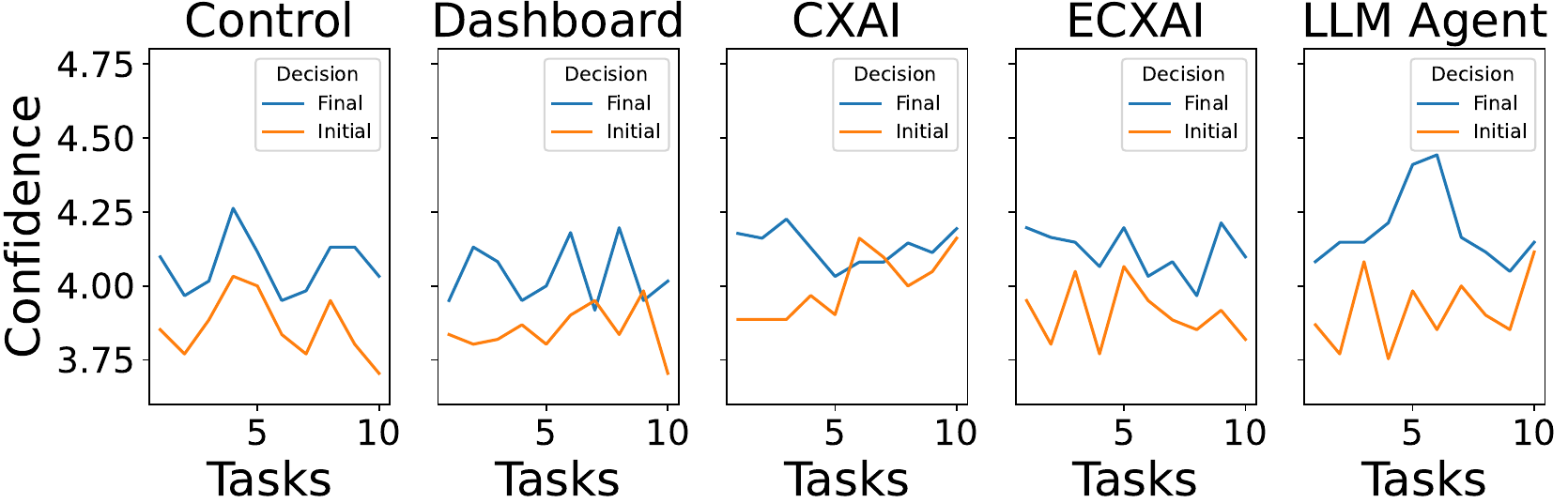}
    \caption{Line plot illustrating the confidence dynamics among users after receiving the AI advice (and explanations). The orange line and blue line illustrate the confidence dynamics before and after receiving AI advice (and explanations), respectively.}
    \label{fig:confidence_dynamics}
    \Description{Four subfigures illustrate the initial user confidence and final user confidence across four experimental conditions. Overall, we found that participants reported a higher confidence after they had access to AI advice. A distinct trend was observed in the \chatxai{} condition, where the reported confidence appears to converge after five tasks. }
\end{figure*}

\subsubsection{Confidence Dynamics}
As shown in Figure~\ref{fig:confidence_dynamics}, we illustrate the confidence dynamics of participants in each condition along with the task order. 
In general, we found that participants reported a higher confidence after being exposed to AI advice and explanations. 
While participants in the \control{} condition, the \dashboard{} condition, and the \chatpersonalized{} condition reported a fluctuating trend of confidence along the task order, participants in the \chatxai{} condition reported a relatively clear ascending trend of confidence both before and after the AI advice (and explanations). 
Participants in the \chatagent{} condition showed a clear upward and then downward trend in their confidence related to their final decisions. 
This suggests that participants in this condition first developed over-confidence in the AI system and then calibrated their confidence.  
Interestingly, we observed that the confidence dynamics of participants in the \chatxai{} condition converge after a few tasks. 
The narrow confidence gap before and after receiving AI advice may indicate that participants in the \chatxai{} condition {calibrate their confidence in the AI advice, which reflects a better understanding of the AI system}.
To compare the confidence across conditions, we conducted ANOVA tests for both initial confidence (average across tasks) and final confidence (average across tasks). 
Although the \chatxai{}, \chatpersonalized{}, and \chatagent{} conditions showed slightly better user confidence on average, we found no significant differences across conditions.

\subsubsection{Further Analysis of User Engagement} 
We measured subjective user engagement reported by each participant in our study using the UES-SF questionnaire~\cite{o2018practical}. The distribution of user engagement across the different experimental conditions was as follows: \control{} ($M = 3.15, SD = 0.72$), \dashboard{} ($M = 3.33, SD = 0.66$), \chatxai{} ($M = 3.20, SD = 0.63$), \chatpersonalized{} ($M = 3.28, SD = 0.67$), \chatagent{} ($M = 3.44, SD = 0.71$). 
While participants in the \chatagent{} condition reported slightly higher engagement with the XAI interface, we found this to be non-significant (based on ANOVA analysis).

\subsubsection{Further Analysis of {Enhanced} Conversation and XAI Usage}\label{sec-xai-usage}
To compare how {enhanced} conversation {(\ie adaptive steering for evaluative decision support and more flexible conversational interactions with LLM agents)} affects user interaction with the conversational interface, we analyzed the usage of the XAI methods.
To compare the usage of each XAI method, we conducted a Kruskal-Wallis H-test for total usage per participant. 
Across all five XAI methods, 
no significant differences in usage frequency were found between the \chatxai{} and \chatpersonalized{} conditions. 
The most obvious difference is that participants in the  \chatxai{} and \chatpersonalized{} conditions used PDP method significantly more frequently: \chatxai{}($M=13.5$), \chatpersonalized{}($M=14.1$), \chatagent{}($M=3.6$). 
Meanwhile, participants in the \chatagent{} condition showed significantly more usage of WhatIF, MACE, and SHAP methods than the \chatxai{} and \chatpersonalized{} conditions.
The reason for such difference in the usage of XAI methods can be caused by the design of the rule-based conversational agent in the \chatxai{} and \chatpersonalized{} conditions. 
In the rule-based conversation agents, all messages are pre-defined, and users see them in a fixed order. 
Such fixed order may have biased user selection of the XAI responses. 
In comparison, the hint questions are randomized in condition  \chatagent{}, and users can also use the free text input to ask anything they prefer. As a result, participants in the \chatagent{} condition may have more flexible access to explore personalized information needs.

\begin{figure}[htbp]
    \centering
    \includegraphics[width=0.48\textwidth]{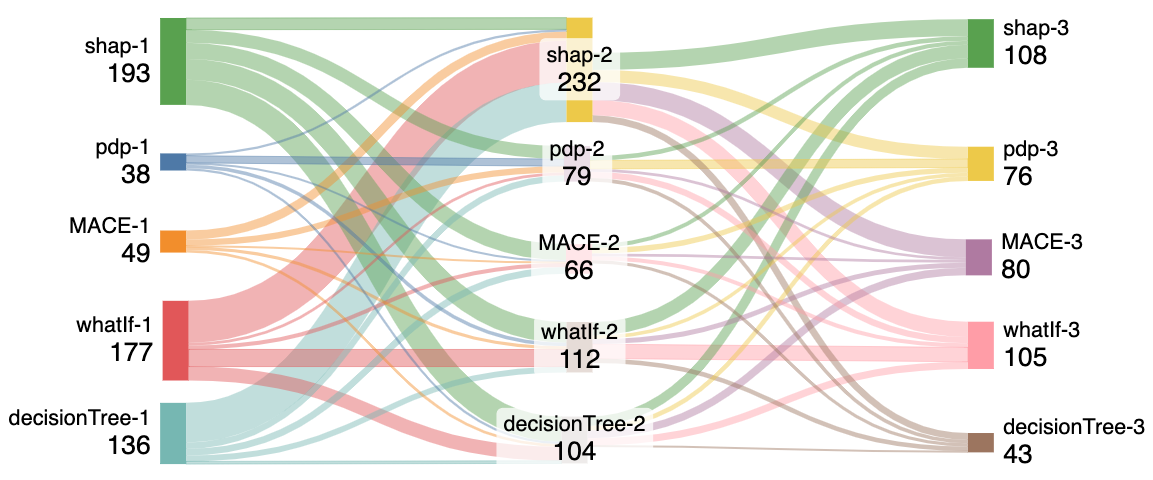}
    \caption{Illustration of the XAI usage used in our study. This Sankey diagram describes the sequence of interactions with XAI methods by users in the  \chatagent{} experimental condition.}
    \label{fig-sankey}
    \Description{{A sankey diagram visualizing the flow of user usage of explainers in the first three steps.}}
\end{figure}

To obtain further insights, we explored the user conversation history in the \chatagent{} condition. 
Among all 61 users in  \chatagent{} , 1,946 user queries are asked in total. Among them, around $40\%$ are based on the hint questions (5 questions we provide to trigger XAI responses, see Table~\ref{tab:xai_conversation}). 
The valid user queries mainly consist of three types of intent: user queries to obtain XAI responses (\eg hint questions and some similar questions), greetings (\eg ``Hi'', ``Thank you''), and opinion-seeking queries to the conversational agent (\eg ``Do you think the loan application is creditworthy?''). 
When meaningless user queries are fired (such as gibberish, random strings or something irrelevant to our task context), the LLM agent-based conversational interface can handle them properly (\eg ``I do not understand this. Please check information related to the current task.'').
To visualize the dynamics of user information needs along with exploring conversation, we adopted the Sankey diagram (Figure~\ref{fig-sankey}) to show the dynamic flow of XAI usage. 
Only a few participants in the \chatagent{} condition asked for more than three XAI responses in each task, so we only considered the first three usages of XAI methods. 
As we can see, after using one XAI method, participants tend to use a different XAI method in the next step, which indicates that most participants explored diverse information needs in the LLM condition \chatagent{}.

\subsection{Additional Discussion}
While no significant results are observed to support the superiority of conversational XAI interface over XAI dashboard, our exploratory analyses revealed the 
potential of conversational XAI interfaces (powered by LLMs) in increasing user exploration of the explanation methods. 
Participants with the conversational XAI interface reported a slightly better perceived user understanding and perceived explanation utility.
As for trust and appropriate reliance, we see that participants showed a slightly higher trust (cf. Section~\ref{sec-h2}), team performance (cf. Section~\ref{sec-h3}), and relatively higher \textit{RAIR} (cf. Table~\ref{tab:res-reliance}). 
We also found that participants with a conversational XAI interface (\chatxai{}, \chatpersonalized{}, and \chatagent{} conditions) did not report a higher user engagement than participants with an XAI dashboard, suggesting that both the interactive interfaces are equally effective in engaging the participants. 

\paratitle{Why {boosted conversations} did not work as expected}. 
In contrast to {our expectation,} 
{boosted conversations (\ie in the \chatpersonalized{} and \chatagent{} conditions)} did not provide further benefits in user understanding, trust, and appropriate reliance. 
According to the confidence dynamics (see Figure~\ref{fig:confidence_dynamics}), 
{enhanced conversation quality in condition \chatagent{}} seems to enlarge the confidence gap between the two stages of decision making (\ie before and after checking AI advice and XAI responses), especially when comparing the \chatagent{} condition with the \chatxai{} condition. 
Although the LLM-powered  condition of \chatagent{} was expected to lead to the most natural and personalized XAI responses among all conditions with XAI interfaces, participants in the  \chatagent{} condition demonstrated the least objective feature understanding, subjective trust, and appropriate reliance. 
Combined with the findings of confidence dynamics, we infer that introducing LLM agents to a conversational XAI interface may amplify the illusion of explanatory depth. 
As a result, participants in the \chatagent{} condition exhibit high over-reliance on the AI system. 
Based on these findings, we argue it would be more important to align the plausibility of XAI responses with the trustworthiness of the AI system rather than solely improving the interactional quality and experiences with the XAI responses. 
This is in line with existing work on plausibility in XAI~\cite{jacovi2020towards}: ``a plausible but unfaithful interpretation may be the worst-case scenario.''
{In comparison, the evaluative conversation enhances user self-reflection of their decision criteria. As a result, participants in condition \chatpersonalized{} indicate a relatively lower \textit{Agreement Fraction} and \textit{RAIR} than condition \chatxai{} (cf. Table~\ref{tab:res-reliance}). 
Thus, we can infer that the evaluative conversation brings about some side effects --- under-reliance on the AI system. 
At the same time, the evaluative conversations fail to facilitate user understanding, calibrate user trust in the AI system, or mitigate over-reliance. 
Further research is required to understand how to provide suitable evaluative decision support in conversational human-AI interactions.}

\paratitle{Potential Bias}. Our study is based on a crowdsourcing setup, which may be affected by cognitive biases introduced in the task design and workflow. 
With the help of the Cognitive Biases Checklist introduced by~\citet{draws2021checklist}, we analyzed potential bias in our study. 
As crowd workers are motivated by monetary compensation, the \textit{self-interest bias} is possible. 
As participants showed a relatively high degree of trust and \textit{Agreement Fraction} with AI advice, 
\textit{Confirmation Bias} may have also affected our results. 
The rule-based conversational agents in the \chatxai{} and \chatpersonalized{} conditions may bias the usage of XAI methods (see Section~\ref{sec-xai-usage}). 
As a result, the participants in the two conditions showed similar usage patterns of XAI methods, which may lead to similar user understanding and reliance patterns.

\end{document}